\documentclass[review, 3p]{elsarticle}

\usepackage{lineno}
\modulolinenumbers[5]

\usepackage{amssymb}
\usepackage{amsmath}
\allowdisplaybreaks
\usepackage{stmaryrd}
\usepackage{esint}

\usepackage[utf8]{inputenc} 
\usepackage[T1]{fontenc}    
\usepackage{amsfonts}       
\usepackage{nicefrac}       
\usepackage{microtype}      
\usepackage{graphicx}
\usepackage{doi}
\usepackage[nomessages]{fp}
\usepackage{nicefrac,xfrac}

\usepackage{placeins}

\numberwithin{equation}{section}
\usepackage{xfrac}
\usepackage{dsfont}
\usepackage{xcolor}
\usepackage{tikz}
\usetikzlibrary{calc}
\usetikzlibrary{decorations.pathreplacing}
\usetikzlibrary{patterns}

\usetikzlibrary{shapes.misc}

\usetikzlibrary{arrows.meta,arrows}
\usetikzlibrary{shapes.geometric}

\tikzset{cross/.style={cross out, draw=black, fill=none, minimum size=2*(#1-\pgflinewidth), inner sep=0pt, outer sep=0pt}, cross/.default={2pt}}

\usepackage{hyperref}
\usepackage{cleveref}

\usepackage{subfigure}
\usepackage{tcolorbox}
\newtcolorbox[list inside=toc]{sectionbox}[1]{colback=white!5!white, colframe=black!75!white, fonttitle=\bfseries, title={\thetcbcounter\ #1}, list entry={\thetcbcounter\quad #1}}
\newtcolorbox{mybox}{colback=red!5!white,colframe=red!75!black}
\def\centerarc[#1](#2)(#3:#4:#5)
    { \draw[#1] ($(#2)+({#5*cos(#3)},{#5*sin(#3)})$) arc (#3:#4:#5); }
    
\usepackage{lineno}

\usepackage{booktabs} 
\usepackage{multirow}
\usepackage{soul} 
\usepackage{microtype}
\usepackage{algorithmic}
\usepackage[lined,boxed,ruled,commentsnumbered]{algorithm2e}

\usepackage{bigints}

\usepackage{multicol}
\usepackage{tabularx}

\newtheorem{thm}{Theorem}
\newtheorem{lem}{Corollary}
\newdefinition{rmk}{Remark}

\journal{Journal of Computational Physics}

\begin{document}
\begin{frontmatter}

\title{Adjoint-based optimization of two-dimensional Stefan problems}

\author[ijlra]{Tomas Fullana \texorpdfstring{\corref{corresponding}}{}}
\cortext[corresponding]{Corresponding author}
\ead{tomas.fullana@sorbonne-universite.fr}

\author[msme]{Vincent Le Chenadec}

\author[ijlra,itv]{Taraneh Sayadi}

\address[ijlra]{Institut Jean le Rond d'Alembert, Sorbonne Universit{\'e}/CNRS, F-75005 Paris, France}
\address[msme]{MSME, Universit{\'e} Gustave Eiffel, F-77454 Marne-la-Vall{\'e}e, France}
\address[itv]{Institute of Combustion Technologies (ITV), RWTH-Aachen University, Aachen, Germany }

\begin{abstract}
	A range of optimization cases of two-dimensional Stefan problems, solved using a tracking-type cost-functional, is presented. A level set method is used to capture the interface between the liquid and solid phases and an immersed boundary (cut cell) method coupled with an implicit time-advancement scheme is employed to solve the heat equation. A conservative implicit-explicit scheme is then used for solving the level set transport equation. The resulting numerical framework is validated with respect to existing analytical solutions of the forward Stefan problem. An adjoint-based algorithm is then employed to efficiently compute the gradient used in the optimisation algorithm (L-BFGS). The algorithm follows a continuous adjoint framework, where adjoint equations are formally derived using shape calculus and transport theorems. A wide range of control objectives are presented, and the results show that using parameterised boundary actuation leads to effective control strategies in order to suppress interfacial instabilities or to maintain a desired crystal shape.  
\end{abstract}


\end{frontmatter}


\section{Introduction}
\label{sec:Introduction}

Stefan problem, named after J. Stefan~\cite{sarler_stefans_1995} due to his substantial contribution to research on moving and free boundaries, model transport and transfer phenomena, in particular solid-liquid phase change in evaporating or chemically reacting flows. Such phenomena govern the interface motion in many engineering related problems such as dendritic solidification~\cite{osher_fronts_1988,Juric1996}, phase transformation in metallic alloys~\cite{Segal1998}, and solid fuel combustion~\cite{Hassan2021}.  

In the applications of interest to this study, the Stefan condition arises from the interaction of liquid and solid phases (both considered incompressible), resulting in a moving liquid-solid interface (freezing or melting front). The speed of the front is directly related to the jump in the conductive heat flux across the interface.  In one dimension, this problem has been studied in depth using various numerical algorithms~\cite{javierre_comparison_2006,Rose1993,Brattkus1992}. In higher dimensions, various methods have been used such as the level set method in Limare \textit{et al.}~\cite{Limare2022} and Osher \textit{et al.}~\cite{osher_fronts_1988} and front-tracking method in Juric \textit{et al.}~\cite{Juric1996}. One of the main challenges associated with modeling such problems in multiple dimensions is due to the unstable dentritic pattern formation~\cite{langer_instabilities_1980,mullins_stability_1964,woods_melting_1992}. In crystal growth, for example, under-cooling triggers an instability mechanism, causing the solid phase of the material to grow into the liquid phase in a finger-like or dendritic fashion, resulting in complex interfacial shapes, which are challenging to predict numerically. In addition, parameters such as anisotropy and surface tension or curvature effect (Gibbs-Thomson) are shown to have a large impact on the dendritic shape of the crystal, which in turn need to be modeled accurately for the simulations to remain predictive. In this work, we present a general framework for tracking and modeling crystal growth in the presence of curvature effects. This algorithm then serves as a vessel to materialise the second, and main, objective of this study which is extracting optimisation strategies to control the resulting solidification process.

The shape of the interface strongly effects the outcome and time-frame of the production process in many industrial applications involving phase change. As a result, while predicting and modeling the resulting solidification process remains at the forefront of many research areas, it is as desirable to extract efficient control strategies to manipulate the motion of the interface, for instance, by tracking a prescribed trajectory. 

Two major types of optimization methods in use today are (i) gradient-based, and (ii) derivative free methods. While an efficient class of generic algorithms (belonging to the class of derivative free methods) based on the surrogate management framework~\cite{Marsden2008} and artificial neural networks~\cite{Pierret2007} have been used for optimization in fluid mechanics, mainly in the area of aerodynamic shape optimization, they could require many function evaluations, for training purposes for example. When detailed simulations of interfacial flows are concerned, each function evaluation commands a full (potentially unsteady) CFD computation, causing gradient-based methods to be at an advantage. Most common methods in extracting the gradient information, on the other hand, are analytical or use finite differences, neither of which are suited to the configuration of interest to this work. Adjoint-based algorithms present a suitable alternative, as they allow the determination of the gradient at a cost comparable to a single function evaluation~\cite{Giles2000}. The use of adjoint methods for design and optimization has been an active area of research which started with the pioneering work of Pironneau~\cite{Pironneau1974} with applications in fluid mechanics, and has been extensively used in aeronautical shape optimization by Jameson and co-workers~\cite{Jameson88,Jameson98}. Ever since these groundbreaking studies, adjoint-based methods have been widely used in fluid mechanics particularly in the areas of aero- and thermo-acoustics~\cite{Juniper2010,Lemke2013}.
More recently flow regimes dominated by nonlinear dynamics, such as separation and mixing have also been analysed using adjoint-based techniques~\cite{Schmidt2013,Rabin2014,Foures2014,Duraisamy2012}. Adjoint-based methods have also been employed for the purpose of sensitivity analysis or control in flows in the presence of large gradients (flames or interfaces)~\cite{Fikl2020,ou_unsteady_2011,Braman2015,Lemke2019}, showing great promise, and therefore are adopted here to carry out the optimisation procedure.

In the context of Stefan problems various control strategies have been employed to track the location of the interface. In a one-dimensional setting, for example, set-valued fixed point equations~\cite{Hoffmann1982} or linear-quadratic defect minimization~\cite{Knabner1985} have been used to control the location of the front. Adjoint-based algorithms have also been applied to a Stefan problem caused by heterogeneous reactions on a surface of a one-dimensional solid particle~\cite{Hassan2021} to extract sensitives with respect to various kinetic parameters. Alternatively, in a two-dimensional setting, adjoint-based algorithms have been utilised previously together with finite element and finite difference approaches to track and control the location of interface by imposing heat flux (or temperature) at the boundary in order to realize the desired interface motion~\cite{Kang1995,Yang1997,Hinze2007}. In particular, Bernauer \& Herzog~\cite{bernauer_optimal_2011}, making use of shape calculus tools, derived the set of adjoint equations to extract control strategies for Stefan problems with a sharp representation of the interface.    

Similar to the approach of~\cite{bernauer_optimal_2011} shape calculus tools have been employed in this work to extract the corresponding adjoint equations. However, contrary to the previous studies, control strategies are extracted here to suppress instabilities of the solidification process (dentritic formation). The addition of curvature effects on the interface (Gibbs-Thomson relation), and the complex shapes encountered during the growth of the crystal require dedicated numerical algorithms capable of modeling both the highly nonlinear forward problem (correct representation of the resulting interface) and the resulting adjoint equations. In addition, while previous studies mostly concentrated on actuation by imposing temperature or heat flux at the boundaries, alternative control strategies using the length of interface or surface tension coefficient are also investigated here, to identify the most relevant and effective (numerical) control strategies in the context of solidification problems.   

The paper is organised as follows. Sec.~\ref{sec:Continuous} describes the equations governing the forward Stefan problem. The details of the numerical dicretization and various schemes used to solve for the temperature field inside the two phases and to track the interface in time are described in Sec.~\ref{sec:discretization}, and validation cases are discussed in Sec.~\ref{sec:Validation}. The adjoint equations are presented in Sec.~\ref{sec:Optimality} and results of the optimisation performed on selected cases are offered and discussed in Sec.~\ref{sec:results}. Summary and perspectives of the work are then discussed in Sec.~\ref{sec:Conclusions}.  

\section{Governing equations in continuous form}
\label{sec:Continuous}

\begin{figure*}[ht]
\begin{center}
    \begin{tikzpicture}[scale = 1.2]
    \draw[very thick, fill = blue!10] (0,0) rectangle (9,4);
    
    \draw[very thick, red, fill = brown!10] (4,2) circle (1.5);
    
    \draw (4,2) node[below] {$\phi(x, t) < 0$};
    \draw (4,2) node[above] {$T_1(x, t)$, $x \in \Omega_1$};
    \draw (1.5,3) node[below] {$\phi(x, t) > 0$};
    \draw (1.5,3) node[above] {$T_2(x, t)$, $x \in \Omega_2$};
    
    \draw (0,2) node[left] {$\partial \Omega$};
    
    \foreach \t [count=\k] in {{$V= -[\nabla T_i]^1_2$},
    {$T_D = T_M - \epsilon_V V - \epsilon_\kappa \kappa$},
    {$\phi(x, t) = 0$},
    {\color{red}$\Gamma$ Interface :}}
  {
   \node at ($(7.3,0.5 + \k/2)$){\t};
  }
    \draw[thick, ->] (5.3,2.75) to (5.9,3.15);
    
    \draw[thick, red, -] (5.5, 2) to (6.4, 2.5);
    
    \draw (5.6,3) node[above] {$n$};
    \end{tikzpicture}
\end{center} 
\caption{Schematic of the two-phase Stefan problem}
\label{fig:Schematic}
\end{figure*}
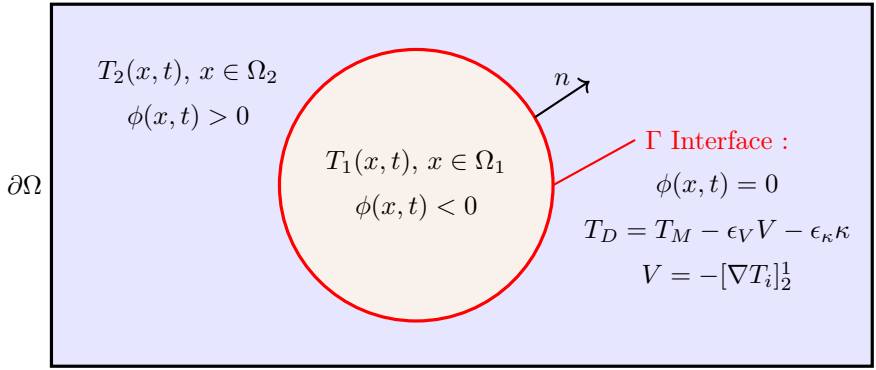

The problem of interest is the Stefan problem in the presence of two immiscible phases with matching densities (one liquid and the other solid). The state of the problem is characterized by the temperature distributions in either phase, as well as the boundary position between them. The computational domain $\Omega$ is partitioned into the time-dependent subdomains $\Omega_1 \left ( t \right )$ and $\Omega_2 \left ( t \right )$ occupied by the liquid ($1$) and solid ($2$) phases, respectively. The external boundary of the domain, denoted $\partial \Omega$, is fixed whereas the interface separating both phases $\Gamma \left ( t \right ) = \overline{\Omega}_1 \left ( t \right ) \cap \overline{\Omega}_2 \left ( t \right )$ evolves in time. A schematic of this configuration is shown in Fig.~\ref{fig:Schematic}.

Let $T_i \colon \left (x, t \right ) \to \mathbb{R} ^ +$ denote the temperature field in either phase ($i = 1, 2$). When the densities $\rho_1$ and $\rho_2$ are equal, convective heat transfers vanish. In addition, when the background pressure is assumed constant, the heat transport equation simplifies to, 
\begin{equation}\label{eq:heat}
\forall i \in \left \llbracket 1, 2 \right \rrbracket, \quad \rho _ i c _ i \frac{\partial T_i}{\partial t} = \nabla \cdot \left ( k _ i \nabla T _ i \right ),  \quad t > 0, \quad x \in \Omega _ i
\end{equation}
where $\rho _ i$, $c _ i$ and $k _ i$ denote the density, the specific heat capacity at constant pressure, and the thermal conductivity, for each phase. Along the interface, energy balance states that
\begin{equation} \label{eq:stefan}
V = -\lambda^{-1} \left ( k_{1} \frac{\partial T_{1}}{\partial n}-k_{2} \frac{\partial T_{2}}{\partial n}\right), \: \: \: x \in \Gamma
\end{equation}
where $\lambda$ denotes the latent heat of solidification and vector $n$ is the outward normal vector at the interface. This jump is taken from phase 1 to phase 2 with $\partial T_i / \partial n$ denoting the normal component of the temperature gradient in phase $i$. Eq.~\ref{eq:stefan} is commonly referred to as the Stefan condition.

In classical Stefan problem, the temperature is set to $T_D(x, t) = T_M$ at the interface where $T_M$ is a constant equal to the melting temperature of the material. For problems involving crystal growth however~\cite{Langer1980}, surface tension effects must be added to the thermodynamic boundary condition by introducing a dependence in the curvature at the front. For that purpose, we use the classical Gibbs-Thomson relation, as defined in Chen \textit{et al.}~\cite{chen_simple_1997}
\begin{equation} \label{eq:gibbs-thomson}
T_D(x, t) = T_M -\epsilon_V V -\epsilon_\kappa \kappa, \quad x \in \Gamma
\end{equation}
where $\kappa$ denotes the curvature at the interface (positive if the center of curvature lies in the solid phase), $V$ the velocity of the interface, $\epsilon_\kappa$ the surface tension coefficient, and $\epsilon_V$ the molecular kinetic coefficient. Unless stated otherwise, both $\epsilon_\kappa$ and $\epsilon_V$ are considered to be constants and the heat capacities, thermal conductivities and latent heat are all set to unity. In addition, to ease the notation, the jump in gradient of temperature is denoted as $[\nabla T]^1_2$ (also defined in Eq.~\ref{eq:jump_gradient}).


When dealing with the numerical approximation of interfacial flows, two classes of methods are commonly used to represent the interface, namely (i) Lagrangian or "front-tracking" methods and (ii) the Eulerian or "front-capturing" methods. The former uses a parameterisation of the interface location (e.g. markers or moving meshes), and has already been used in Stefan problems~\cite{juric_front-tracking_1996} but has not been adopted in this work due to the inherent difficulty of deriving the continuous adjoint equations with such methods (see Sec.~\ref{sec:Optimality}). The latter can broadly speaking be divided into two categories: Volume-Of-Fluid (VOF) and Level Set methods. In adjoint-based optimization, the VOF method~\cite{Fikl2020} may lead to numerical complications due to the piece-wise linear reconstruction of the interface. On the other hand, the Level Set method, where the interface is implicitly defined as a continuous function, has been proven to be well suited in continous adjoint-based optimization, specifically for Stefan problems~\cite{bernauer_optimal_2011}. Moreover, this method has been shown to accurately reproduce dendritic pattern formation~\cite{chen_simple_1997, Limare2022}. An implicit signed distance strategy has therefore been used here to represent the interface.

A level set function $\phi \colon (x, t) \to \mathbb{R}$ is constructed, such that, at any time $t$, the front is equal to the zero level set of the function 
\begin{equation} \label{eq:LSinterface}
\Gamma(t) =  \{ x \in \Omega : \phi(x, t) = 0 \} .
\end{equation}
The level set function is initially set to the signed distance function, with $d$ the distance to the front, such that
\begin{equation} \label{eq:LSsigneddistance}
\phi(x, 0) = \left\{
\begin{array}{rr} 
+d, \quad x \in \Omega_2, \\ 
0, \quad x \in \Gamma, \\ 
-d, \quad x \in \Omega_1.
\end{array}\right .
\end{equation}
The level set function is advected with a speed function $F$ which is extended from the normal speed $V$ at the front. The equation of motion of $\phi$ is given by the level set advection equation 
\begin{equation} \label{eq:LSadvection}
\dfrac{\partial \phi}{\partial t} + F |\nabla \phi| = 0.
\end{equation}
Equation (\ref{eq:LSadvection}) will move $\phi$ with the correct speed such that $\Gamma$ remains the zero level set of $\phi$. The field $\phi$ is also used to define the normal vector $n$ by
\begin{equation} \label{eq:LSnormal}
n=\dfrac{\nabla \phi(x, t)}{|\nabla \phi(x, t)|}, \: \: \: x \in \Gamma \left ( t \right )
\end{equation}
and the curvature $\kappa$ by
\begin{equation} \label{eq:LSkappa}
\kappa=\nabla \cdot \left(\dfrac{\nabla \phi(x, t)}{|\nabla \phi(x, t)|}\right), \: \: \: x \in \Gamma \left ( t \right ).
\end{equation}

In two dimensions, the curvature $\kappa$ at the front is computed in a non-conservative form using
\begin{equation}\label{eq:meancurvature}
\kappa=\frac{\left(\phi_{y}^{2} \phi_{x x}-2 \phi_{x} \phi_{y} \phi_{x y}+\phi_{x}^{2} \phi_{y y}\right)}{\left(\phi_{x}^{2}+\phi_{y}^{2}\right)^{3 / 2}}.
\end{equation} 

Considering the above mentioned equations and relations, the following generic forward two-phase Stefan problem (\ref{FP}) can be recast in the level set framework given below, which is later used in the adjoint derivation. 
\begin{center}
Find a function $T : \Omega \times[0, t_{f}] \rightarrow \mathbb{R}$ and a function $\phi : \Omega \times[0, t_{f}] \rightarrow \mathbb{R}$ such that:
\[\arraycolsep=2pt\def\arraystretch{2.2}
\label{FP} \tag{FP}\left\{
\begin{array}{rcll} \dfrac{\partial T_{1}}{\partial t} &=& \Delta T_{1} & \: \text{in} \: \Omega_1(t) \\
\dfrac{\partial T_{2}}{\partial t} &=& \Delta T_{2} & \: \text{in} \: \Omega_2(t) \\ 
T(x,0) &=& T_0 \left ( x \right ) & \: \text{in}  \: \Omega \\
\displaystyle \dfrac{\partial T(x,t)}{\partial n} &=& u \left ( x, t \right ) & \: \text{on} \: \partial \Omega\\
T(x,t) &=& T_M -\epsilon_V V -\epsilon_\kappa \kappa & \: \text{on} \: \Gamma(t) \\
\dfrac{\partial \phi}{\partial t} &=& -[\nabla T_i]^1_2 \cdot \nabla \phi & \: \text{on} \: \Gamma(t) \\
\phi(x,0) &=& \phi_0 \left ( x \right ) & \: \text{in} \: \Omega 
\end{array}
\right.
\]
\end{center}
Here, $T _ i$ denote the restrictions of $T$ to $\Omega _ i \left ( t \right )$, and $T_0$ and $\phi_0$ are the initial conditions at $t = 0$ of the temperature field and the level set function, respectively. The velocity coefficient in the Dirichlet boundary condition at the interface $\epsilon_V$ is set to zero.

\section{Numerical framework}
\label{sec:discretization}

In this section, the methods used to solve both the forward and the adjoint Stefan problems are discussed. The solution procedure is split into six main steps: (i) solution of the heat equation in each phase, (ii) computation of the Stefan condition, (iii) extension of the velocity field normal to the interface, (iv) propagation of the level-set function, (v) reinitialisation of the level-set function (so as to satisfy the signed distance property), and finally (vi) handling of the dead and fresh cells. Each of these steps and their corresponding convergence properties are presented in the following subsections and summarised in Algorithm~\ref{alg:methodoutline}. 


\begin{algorithm*}[ht]

	\SetKwData{Left}{left}
	\SetKwData{This}{this}
	\SetKwData{Up}{up}
	\SetKwInOut{Input}{input}\SetKwInOut{Output}{output}
	\Input{$T_0, \: T_D, \: \lambda, \: \kappa_1, \: \kappa_2, \: c_1, \: c_2$}
	\Output{$\phi, \: T$}

	\BlankLine \emph{Initialize the level set function }$\phi_0$ \emph{and the temperature field} $T_0$\\
	
	\Repeat{final time $t \leftarrow t_{f}$}{
	
	\BlankLine \textbf{1.} \emph{Update the temperature field : }$\tilde{T}_n \leftarrow T_{n-1}, \: T_D$ \hfill (Sec.~\ref{subsec:Temperature})\\
	\BlankLine \textbf{2.} \emph{Compute the Stefan condition : } $V \leftarrow \lambda, \: \kappa_1, \: \kappa_2, \: c_1, \: c_2, \: [\nabla T]^{1}_{2} $ \hfill (Sec.~\ref{subsec:Stefan})\\
	\BlankLine \textbf{3.} \emph{Extend the velocity field : } $F \leftarrow V$ \hfill (Sec.~\ref{subsec:Extension})\\
	\BlankLine \textbf{4.} \emph{Update the level set function :} $\phi_{n} \leftarrow \phi_{n-1}, F$ \hfill (Sec.~\ref{subsec:IIOE})\\
	\BlankLine \textbf{5.} \emph{Reinitialize the level set function :} $\tilde{\phi}_{n} \leftarrow \phi_{n}$ \hfill (Sec.~\ref{subsec:Reinitialisation})\\
	\BlankLine \textbf{6.} \emph{Clean or Initialize dead or fresh cells :} $T_n \leftarrow \tilde{T}_n$ \hfill (Sec.~\ref{subsec:Deadfresh})\\
	\BlankLine
	}
	\caption{Outline of the method}\label{alg:methodoutline}
\end{algorithm*}

\subsection{Temperature update} 
\label{subsec:Temperature}

The heat equation (Eq.~\ref{eq:heat}) is discretized on the subdomains $\Omega _ 1 \left ( t \right )$ and $\Omega _ 2 \left ( t \right )$ using a Cartesian grid (identical to the one used to solve the level set advection equation, described in detail in Sec.~\ref{subsec:IIOE}). Since the interface $\Gamma \left ( t \right )$ might not align with either of the grid axes, the discretization of the Laplacian operator is carried using a recently proposed Cut Cell Method~\cite{FullanaCutCell}. The underlying idea is to modify the standard centered difference formulas for approximating the second derivatives to impose the proper boundary condition (Eq.~\ref{eq:gibbs-thomson}) on the interface. Here, the procedure for imposing the Dirichlet boundary condition at the interface is briefly describe (for details regarding a more general implementation of the operators, including their modification to account for Neumann boundary conditions on $\partial \Omega$, the reader is referred to~\cite{FullanaCutCell}).  

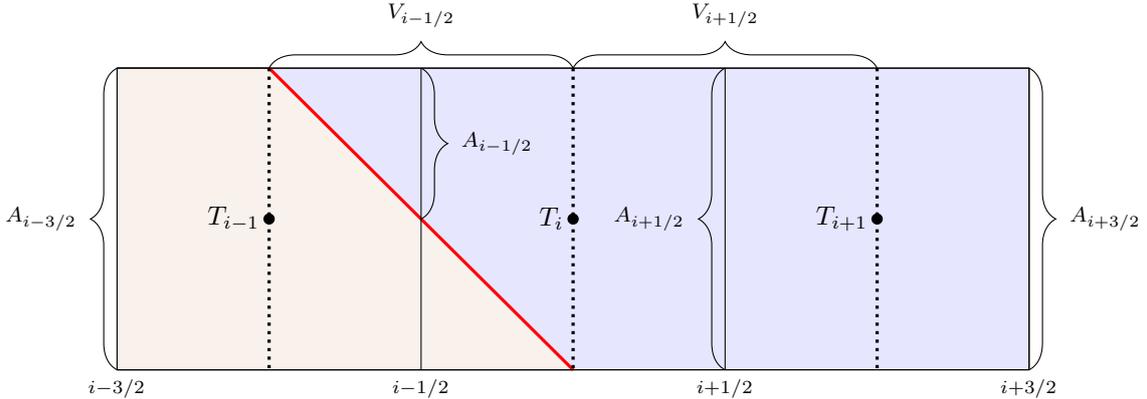
\begin{figure*}[ht!]
\begin{center}
    \begin{tikzpicture}[scale = 2.0]
    
    \draw[fill = blue!10] (-3,-1) rectangle (3,1);
    \draw[fill = brown!10] (-3,-1) -- (0,-1) -- (-2,1) -- (-3,1);
    
    \node at (-2,0) [circle,fill,inner sep=1.5pt]{};
    \node at (0,0) [circle,fill,inner sep=1.5pt]{};
    \node at (2,0) [circle,fill,inner sep=1.5pt]{};
    
    \draw (-2,0) node[left] {$T_{i-1}$};
    \draw (0,0) node[left] {$T_{i}$};
    \draw (2,0) node[left] {$T_{i+1}$};
    
    \draw [red, very thick] (-2,1) to (0,-1);
    \foreach \ii in {1,-1}
    	{\draw (-3,\ii) to (3,\ii);}
    \foreach \ii in {-3,1,-1,3}
    	{\draw (\ii,-1) to (\ii,1);}
    
    \draw [decorate,decoration={brace,amplitude=10pt,mirror}]
(-1,0) -- (-1,1) node [midway,xshift=1cm] {\footnotesize$A_{i-1/2}$};
\draw [decorate,decoration={brace,amplitude=10pt}]
(1,-1) -- (1,1) node [midway,xshift=-1cm] {\footnotesize$A_{i+1/2}$};
\draw [decorate,decoration={brace,amplitude=10pt}]
(-3,-1) -- (-3,1) node [midway,xshift=-1cm] {\footnotesize$A_{i-3/2}$};

\draw [decorate,decoration={brace,mirror,amplitude=10pt}]
(3,-1) -- (3,1) node [midway,xshift=1cm] {\footnotesize$A_{i+3/2}$};

\draw [decorate,decoration={brace,amplitude=10pt}]
(-2,1) -- (0,1) node [midway,yshift=0.7cm] {\footnotesize$V_{i-1/2}$};

\draw [decorate,decoration={brace,amplitude=10pt}]
(0,1) -- (2,1) node [midway,yshift=0.7cm] {\footnotesize$V_{i+1/2}$};

\draw [dotted, very thick] (-2,1) to (-2,-1);
\draw [dotted, very thick] (0,1) to (0,-1);
\draw [dotted, very thick] (2,1) to (2,-1);

\draw (-3, -1) node[below] {$_{i-3/2}$};
\draw (-1, -1) node[below] {$_{i-1/2}$};
\draw (1, -1) node[below] {$_{i+1/2}$};
\draw (3, -1) node[below] {$_{i+3/2}$};

    
    \end{tikzpicture}
\end{center}   

\caption{Example of face capacities and staggered volumes necessary to compute the discrete homogenenous and inhomogeneous gradient operators in the positive $x$ direction (Equations~\ref{eq:homogeneous} and ~\ref{eq:inhomogeneous}). }
\label{fig:cutcell}
\end{figure*}

Following the Cut Cell strategy, the interface position is used to compute the areas (volume, respectively) wetted by each phase in each face (cell, respectively), henceforth referred to as face-centered surface (denoted $A _ {i \pm 1/2, j}$ and $A _ {i, j \pm 1/2}$) and cell-centered volume (denoted $V _ {i, j}$) capacities. A piece-wise bi-quadratic interpolation of the level-set function values is used to determine the crossing points of the interface with the Cartesian grid, and supplied to the marching squares algorithm that infers the face and volume capacities of partially-filled faces and cells (Fig.~\ref{fig:cutcell}). The construction of the operators is briefly described below for the $x$ contribution to the Laplacian, for Dirichlet boundary conditions.

In continuous form, the $x$ component of the temperature gradient may be integrated over a volume $\Omega$ filled by any given phase, and transformed according to Stokes' theorem to yield,
\begin{equation}\label{eq:Stokes}
\int _ \Omega \frac{\partial T}{\partial x} \mathrm{d} V = \oint _ {\partial \Omega} T \mathbf{e} _ x \cdot \mathrm{d} \mathbf{S}
\end{equation}
where $\mathrm{d} \mathbf{S}$ denotes the outward-pointing surface element and $\mathbf{e} _ x$ the unit vector along the $x$ direction. If $\Omega = \Omega _ {i + 1/2,j}$ consists of the intersection of a phase domain and a $x$ face-centered computational cell $\left ( i + 1/2, j \right )$, its contour may be decomposed as $\partial \Omega = S _ {i, j} ^ - \cup S _ {i + 1, j} ^ + \cup S _ {i + 1/2, j-1/2} ^ - \cup S _ {i + 1/2, j+1/2} ^ + \cup \left ( \Omega \cap \Gamma \right )$, where the $\pm$ subscript denotes the direction the mesh-aligned surfaces point into. Inserted in Eq.~\ref{eq:Stokes}, this decomposition leads to,
\begin{equation}\label{eq:gradx}
g _ {i + 1/2, j} \equiv \int _ {\Omega _ {i + 1/2, j}} \frac{\partial T}{\partial x} \mathrm{d} V = \underbrace{\int _ {S _ {i + 1, j}} T \mathrm{d} S - \int _ {S _ {i, j}} T \mathrm{d} S} _ {\mathrm{fluid} \quad (g ^ \Omega _ {i + 1/2, j})} + \underbrace{\int _ {\Omega _ {i + 1/2, j} \cap \Gamma} T \mathbf{e} _ x \cdot \mathrm{d} \mathbf{S}} _ {\mathrm{boundary} \quad (g ^ \Gamma _ {i + 1/2, j})}   
\end{equation}
where the right-hand side is decomposed into fluid and boundary contributions, respectively approximated as,
\begin{equation}\label{eq:homogeneous}
g ^ \Omega _ {i+1/2,j} = \frac{A_{i+3/2,j}+A_{i+1/2,j}}{2}T_{i+1,j} - \frac{A_{i+1/2,j}+A_{i-1/2,j}}{2}T_{i,j},
\end{equation}
and,
\begin{equation}\label{eq:inhomogeneous}
g^{\Gamma}_{i+1/2,j} = \left ( \frac{A_{i+3/2,j}-A_{i+1/2,j}}{2} + \frac{A_{i+1/2,j}-A_{i-1/2,j}}{2} \right ) T_D.
\end{equation}

To preserve symmetry, the discrete divergence is set to the negative transpose of the (fluid) gradient operator, which yields the following approximation of the temperature Laplacian,
\begin{equation}
\int _ {\Omega _ {i, j}} \Delta T \mathrm{d} V \simeq 
\frac{A_{i+1/2,j}+A_{i-1/2,j}}{2} \left ( \frac{g_{i+1/2,j}}{V_{i+1/2,j}} - \frac{g_{i-1/2,j}}{V_{i-1/2,j}} \right ) + \frac{A_{i,j+1/2}+A_{i,j-1/2}}{2} \left ( \frac{g_{i,j+1/2}}{V_{i,j+1/2}} - \frac{g_{i,j-1/2}}{V_{i,j-1/2}} \right )
\end{equation}
where the staggered volumes ($V_{i \pm 1/2,j}$ and $V_{i, j \pm 1/2}$) are interpolated from $V _ {i,j}$.

The values computed in the neighbouring points are projected back to the interface centroid in the partial cell using a bi-quadratic interpolation. Using the described discretisation and boundary conditions, the heat equation is then propagated in time using a Crank-Nicolson scheme. 

\begin{figure*}[ht!]
\begin{center}
	\includegraphics[width=0.7\textwidth]{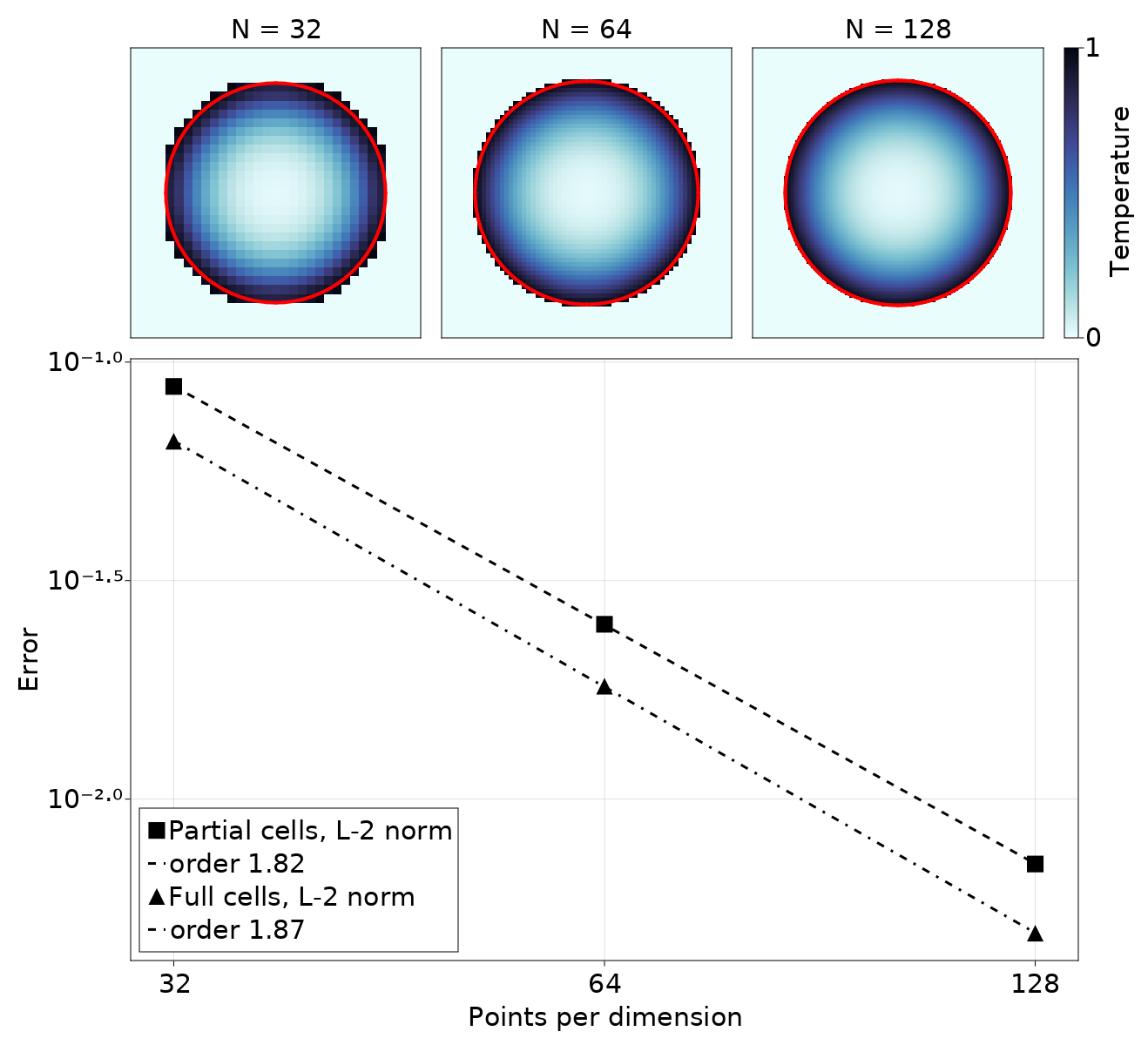}
\caption{Convergence study of the Cut Cell method in a stationary geometry with a Dirichlet boundary condition $T_D = 1$ imposed at the interface. The top figures show the position of the interface in red and the temperature field at final time for different resolutions. The bottom figure shows the convergence rate of the method in partial cells and in full cells.}
\label{fig:cutcell_validation}
\end{center}
\end{figure*}

The implementation of the Cut Cell method coupled with a Crank-Nicolson scheme is validated in the following stationary setup. A solid circle of radius $R = 0.75$ is initialized in a $2\times 2$ domain. The initial temperature field is set to zero and a Dirichlet boundary condition is imposed at the interface $T_D = 1$. We solve only for the phase inside of the circle until a final time $t_f = 0.03125$. The simulation is performed for different resolutions N = 32, 64, 128 with a fixed CFL = 0.5 corresponding to 16, 64, 256 iterations, respectively. In Fig.~\ref{fig:cutcell_validation}, the final temperature fields and the error in partial and full cells are shown. In stationary geometry, the Cut Cell method coupled with a Crank-Nicolson scheme shows an order of convergence slightly below 2 for both the partial and full cells.


\subsection{Stefan condition}
\label{subsec:Stefan}

\begin{figure*}
    \centering
    \subfigure[]
    {
    \begin{tikzpicture}[scale=2.3]
    
            \draw[fill = blue!10] (0,0) rectangle (3,3);
            \draw[fill = brown!10] (0.7, 3) -- (0.3, 0) -- (0,0) -- (0,3);
            
            \foreach \ii in {0, 1, 2, 3}
                	{\draw [very thick] (0, \ii) to (3,\ii);}
            \foreach \ii in {0, 1, 2, 3}
                	{\draw [very thick] (\ii, 0) to (\ii, 3);}
            
            \draw [very thick, red] (0.7, 3) to (0.3, 0);
            
            \foreach \ii in {0.5, 1.5, 2.5}
            	{\FPeval{\result}{clip(0.5 + \ii)}%
            	\node at (1.5,\ii) [circle,fill,inner sep=2.5pt]{};
            	\node at (2.5,\ii) [circle,fill,inner sep=2.5pt]{};
            	\draw (1.5, \ii) node[right, xshift=5pt] {$T^{\result}_A$};
            	\draw (2.5, \ii) node[right, xshift=5pt] {$T^{\result}_B$};}
            	
            \draw [very thick, dotted] (2.5, 2.5) to (2.5, 0.5);
            \draw [very thick, dotted] (1.5, 2.5) to (1.5, 0.5);
            \draw [very thick] (0.5, 1.5) to (2.5, 1.1);
            
            \node at (0.5, 1.5) [red, rectangle,fill,inner sep=3.5pt]{};
            \node at (1.5, 1.3) [violet,fill,inner sep=3.5pt]{};
            \node at (2.5, 1.1) [blue,fill,inner sep=3.5pt]{};
            
            \draw [-{Stealth[length=3mm, width=2mm]}, very thick, black] (0.55, 1.75) to (0.8, 1.7);
            
            \draw (0.675, 1.725) node[above, yshift=2pt] {$n$};
            
            \draw (0.5, 1.5) node[red, left, yshift=10.5pt] {\large{$T_D$}};
            \draw (1.5, 1.3) node[violet, left, yshift=-10.7pt] {\large{$T^*_A$}};
            \draw (2.5, 1.1) node[blue, above, xshift=15.5pt] {\large{$T^*_B$}};
            
            \draw (2, 1.2) node[blue, right, yshift=5pt] {\large{$d_B$}};
            
            \draw (1, 1.4) node[violet,left, yshift=-5pt] {\large{$d_A$}};
            
            \draw (0.3, 2.8) node[] {\large{Solid}};
            \draw (1.5, 2.8) node[] {\large{Liquid}};
            
            \end{tikzpicture}
    \label{fig:Stefan_condition_a}
    }
    \subfigure[]
    {  
    \includegraphics[width=0.5\textwidth]{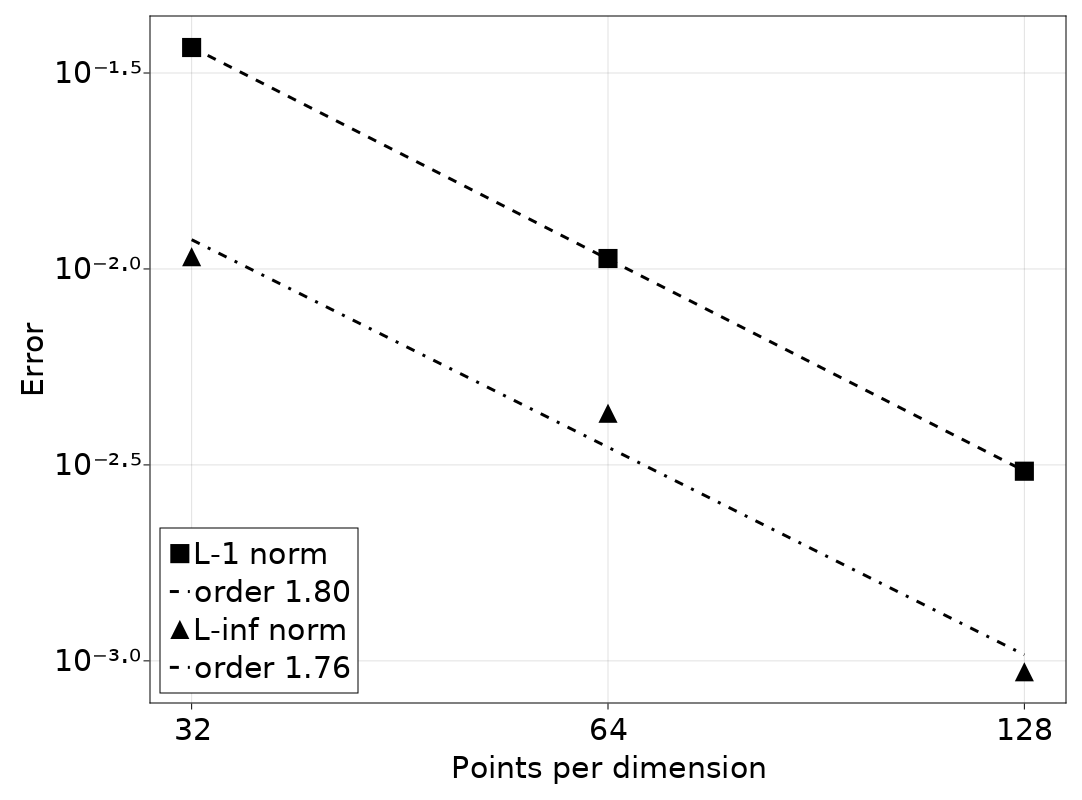}
    \label{fig:Stefan_condition_b}
    }
    \caption{(a) Schematic of gradient calculation. The Dirichlet value $T_d$ is imposed at the interface centroid in the partial cell and the temperatures $T_A$ and $T_B$ are determined via a quadratic interpolation from the neighboring 3 points in the vertical direction (dotted lines). (b) Convergence of the gradient computation.}
\end{figure*}
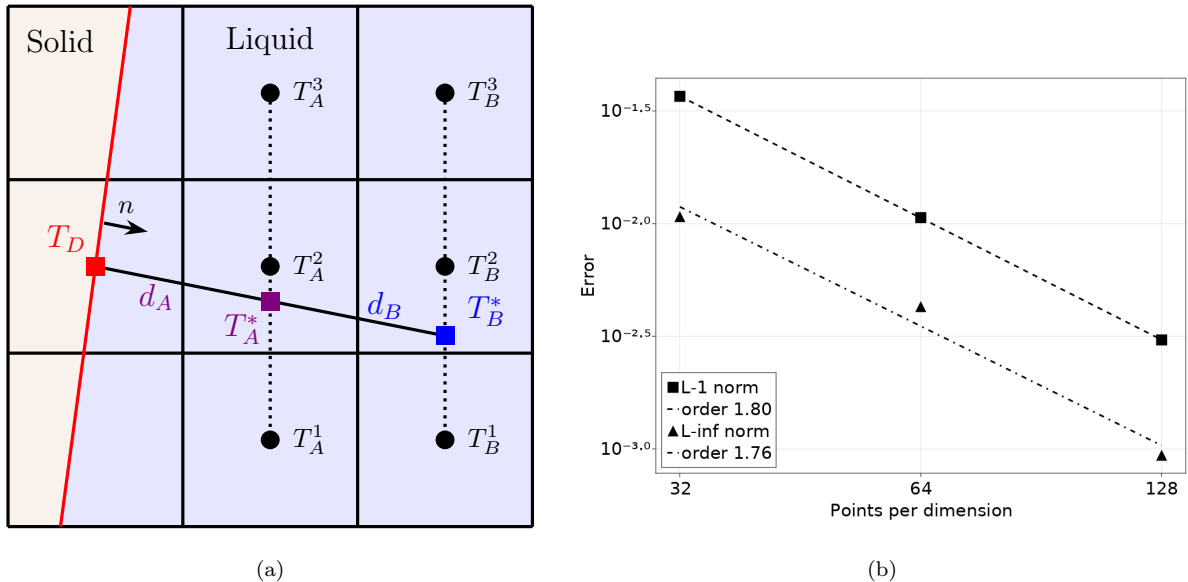

As stated previously, the motion of the interface is solely dependent on the jump in the temperature gradient. It is therefore important to compute the normal gradient in temperature of each phase accurately, and to this end, the Johansen-Colella method~\cite{johansen_cartesian_1998} is used
\begin{equation}\label{eq:Johansen_Colella}
\nabla T |_{\Gamma} = \dfrac{1}{d_B - d_A} \left( \dfrac{d_B}{d_A} (T_D - T^*_A) - \dfrac{d_A}{d_B} (T_D - T^*_B) \right)
\end{equation}
with $T_D$ the dirichlet value imposed at the interface, $T^*_A$ and $T^*_B$ the interpolated values of the temperature field on points $A$ and $B$ respectively, and $d_A$ and $d_B$ the distances between the interface centroid to $A$ and $B$ respectively. The procedure for computing the gradient in one phase is as follows: (i) a shifted $3\times 3$ stencil is chosen, as shown in Fig.~\ref{fig:Stefan_condition_a}, (ii) a line from the interface centroid is cast in the normal direction $n$, (iii) the crossing points $A$ and $B$ of this line and the vertical (or horizontal, depending on the normal orientation) segments of the neighboring 3 points are identified and the distances $d_A$ and $d_B$ are computed, and finally (iv) the values $T^*_A$ and $T^*_B$ are interpolated using $T^1_A, T^2_A, T^2_A$ and $T^1_B, T^2_B, T^2_B$, respectively.
Once the normal gradient is computed in each phase, using Eq.~\ref{eq:Johansen_Colella}, the jump is computed as
\begin{equation}\label{eq:jump_gradient}
[\nabla T]^1_2 = \nabla T_1 |_{\Gamma} - \nabla T_2 |_{\Gamma}.
\end{equation}
The discrete velocities of the front in the partial cells are initialized with this jump and will be used as the boundary condition in the velocity extension procedure. To validate the method within our Cut Cell framework, we consider a stationary circle of radius $R = 0.5$ in a $1 \times 1$ domain and we initialize the temperature field with a similarity solution of the heat equation
\begin{equation}\label{eq:similarity}
    T(r) = \left\{
\begin{array}{lr} 
T_\infty \left( 1 - \dfrac{F(r)}{F(R)} \right), & r > R \\
0, & r < R
\end{array}\right.
\end{equation}
with $T_\infty = -0.5$ a given under-cooling temperature, and $F(r) = \operatorname{E}_1 (1/4 \: r^2)$ where $\operatorname{E}_1(t) = \int_x^\infty \frac{\operatorname{e}^{-t}}{t} dt$. The discrete velocities at the front are computed for different resolutions N = 32, 64, 128, 256 and the resolution $256 \times 256$ is used as the reference solution for the convergence study. Fig.~\ref{fig:Stefan_condition_b} shows that the gradient extraction procedure coupled with the interface location method results in close to second-order accuracy in both $L_2$ and $L_\infty$ norms. 

\subsection{Velocity extension}
\label{subsec:Extension}

\begin{figure*}[ht!]
        \centering
        \includegraphics[width=1\textwidth]{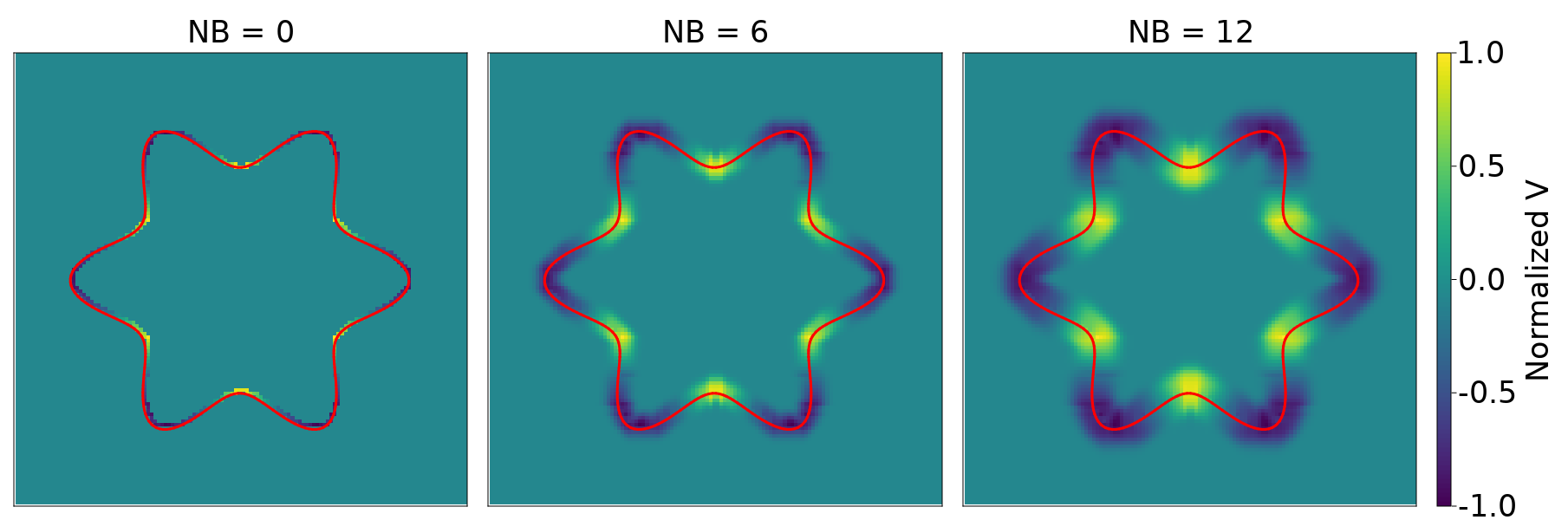}
        \caption{Velocity fields of a crystal-shape geometry. The red curves represent the interface location and the velocity fields correspond, from left to right, to narrow bands widths of 0, 6 and 12.}
        \label{fig:narrowband}
\end{figure*}

As highlighted in Sec.~\ref{sec:Continuous}, the front velocity needs to be extended away from the interface. The most natural procedure is to let $V$ be a constant along the lines normal to $\Gamma$. To achieve this, the method described in Peng \textit{et al.} is adopted here~\cite{peng_pde-based_1999}. Using this approach, the velocity is extended in the normal direction by solving the following hyperbolic partial differential equation
\begin{equation}\label{eq:velo_extension}
\left\{\begin{array}{rr}
\dfrac{\partial F}{\partial \tau}+S(\phi) \dfrac{\nabla \phi}{|\nabla \phi|} \cdot \nabla F=0&\text{in} \: \Omega \\
F(x, 0)=V & \text{on} \: \Gamma
\end{array}
\right.
\end{equation} 
where $F$ is the extended velocity field equal to $V$ at the front, $\tau$ denotes a pseudo-time and $S(\phi)$ is the signature function 
\begin{equation}\label{eq:signature}
S(\phi)=\left\{\begin{array}{ll}
-1 & \text { if } \phi<0 \\
0 & \text { if } \phi=0 \\
+1 & \text { if } \phi>0
\end{array}\right.
\end{equation} 
Eq.~\ref{eq:velo_extension} is then discretized using a first order upwind scheme and integrated in time by a forward Euler method until steady state. Taking $n$ as the normal vector defined as
\begin{equation}\label{eq:normal_vector}
n = (n_x, n_y)  = \left(\phi_x / \sqrt{(\phi_x^2 + \phi_y^2)}, \phi_y / \sqrt{(\phi_x^2 + \phi_y^2)} \right),
\end{equation}
the discretisation leads to
\begin{equation}\label{eq:discrete_velo}
\begin{aligned}
F_{i j}^{n+1}=& F_{i j}^{n}-\Delta \tau\left(\left(S_{i j} n_{i j}^{x}\right)^{+} \displaystyle \dfrac{F_{i j}-F_{i-1 j}}{h}+\left(S_{i j} n_{i j}^{x}\right)^{-} \dfrac{F_{i+1 j}-F_{i j}}{h}\right.\\
&\left.+\left(S_{i j} n_{i j}^{y}\right)^+\displaystyle \dfrac{F_{i j}-F_{i j-1}}{h}+\left(S_{i j} n_{i j}^{y}\right)^-\dfrac{F_{i j+1}-F_{i j}}{h}\right)
\end{aligned}
\end{equation}
where $h$ is the uniform grid spacing, $(x)^+ = \operatorname{max}(0,x)$ and $(x)^- = \operatorname{min}(0,x)$, and the time step $\Delta \tau$ is chosen so that $\Delta \tau / h = 0.45$. The pseudo-time spawn in the velocity extension procedure is purely ficticious and the number of iterations in Eq.~\ref{eq:discrete_velo} corresponds to the width of the narrow-band (NB) around the 0-level set where the velocities are initialized. Fig.~\ref{fig:narrowband} shows an example of an initial velocity field for different narrow-band widths, after one iteration of the heat equation, where the temperature at the interface $T_D = \epsilon_\kappa \kappa$ depends only on the curvature. The computed discrete velocities are positive in the kinks and negative in the tips driving the initial crystal towards a circular shape. 

\subsection{Level Set advection equation} 
\label{subsec:IIOE}

A semi-implicit scheme described in Mikula \textit{et al.}~\cite{mikula_inflow-implicitoutflow-explicit_2014, mikula_new_2010} is used here to solve the level set advection equation (Eq.~\ref{eq:LSadvection}). This scheme is briefly described below (for details of the scheme the reader is referred to \cite{mikula_inflow-implicitoutflow-explicit_2014}). Using this method, the equation is written in an alternative form
\begin{equation}\label{eq:advection0}
    \dfrac{\partial \phi}{\partial t} + F \dfrac{\nabla \phi}{|\nabla \phi|} \cdot \nabla \phi = 0
\end{equation}
which is then divided into conservative and non-conservative terms
\begin{equation}\label{eq:advection1}
    \dfrac{\partial \phi}{\partial t} + \underbrace{\nabla \cdot \left( F \phi \dfrac{\nabla \phi}{|\nabla \phi|} \right)}_{A} - \underbrace{\phi \nabla \cdot \left( F \dfrac{\nabla \phi}{|\nabla \phi|} \right)}_{B} = 0,
\end{equation} 
resulting in a second order partial differential equation akin to a weighted diffusion equation. The first term ($A$) has a diffusion coefficient $F \phi$ that depends on the solution and represents a nonlinear curvature flow whereas in the second term ($B$) the solution is multiplied by the curvature of its level-sets. The main idea behind Mikula \textit{et al.}~\cite{mikula_inflow-implicitoutflow-explicit_2014} scheme is to distinguish two cases: if the product $F \phi$ is negative (positive, respectively) then $A$ represents a forward (backward, respectively) diffusion and $B$ represents a backward (forward, respectively) diffusion. The forward diffusion is treated implicitly while the backward diffusion is treated explicitly leading to a semi-implicit scheme with a diffusive CFL number.

\begin{figure*}[ht!]
        \centering
        
        \includegraphics[width=0.8\textwidth]{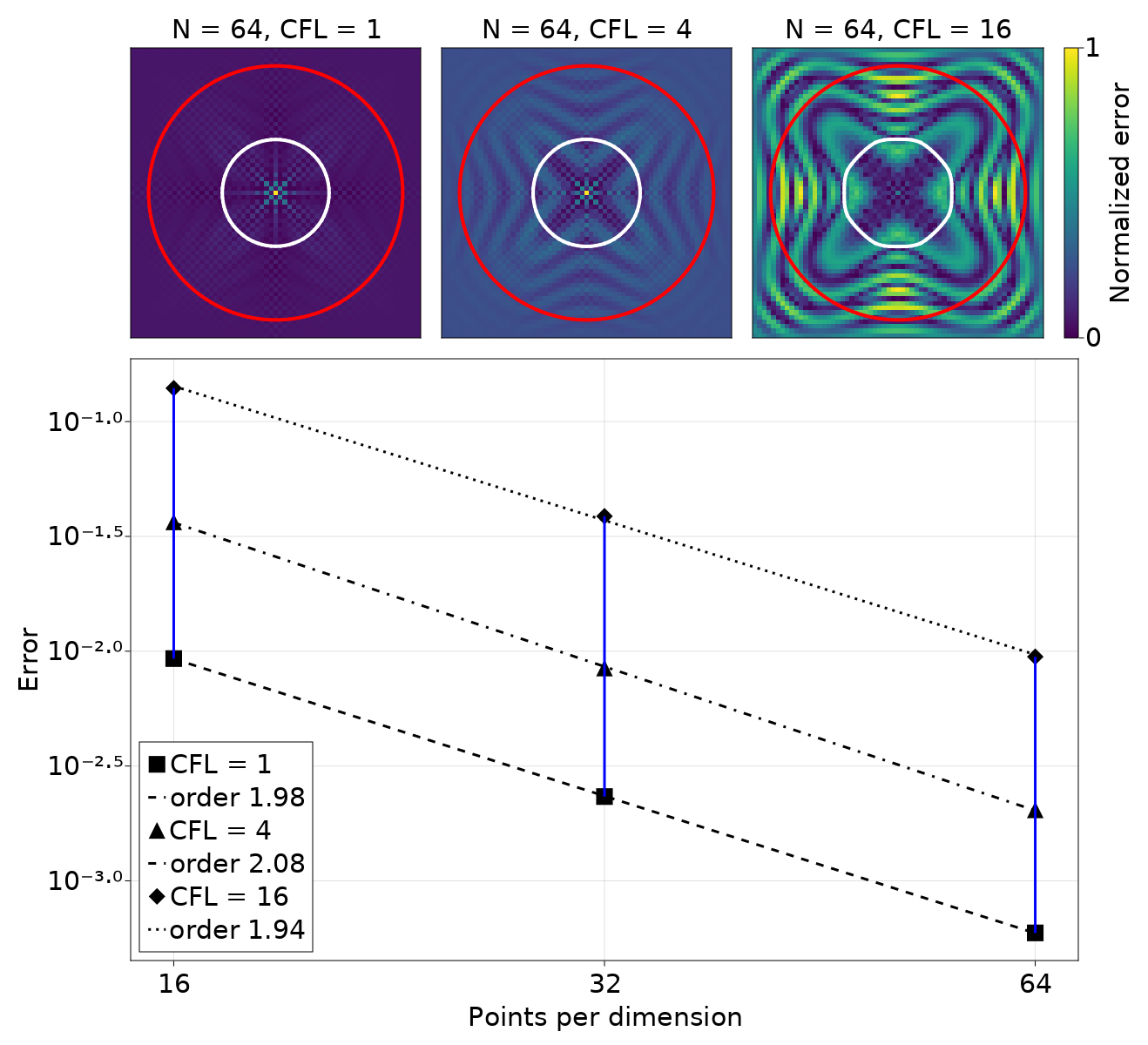}
        \caption{The top figures represent the normalized error field for different CFL = 1, 4, 16 and fixed N = 64. The red (white, respectively) curve represents the initial (final, respectively) 0-level set. In the bottom figure the error in $L_2$ norm is plotted for different resolutions and CFL numbers. The vertical blue lines correspond to a fixed number of points for varying CFL numbers.}
        \label{fig:CFL}
\end{figure*}

This scheme allows us to relax the CFL condition, usually present in most of level set methods. To check the robustness of the method for a $\nicefrac{\Delta t}{h^2}$ ratio exceeding the usual CFL condition, we consider an initial level set function $\phi_0$ in a $1 \times 1$ domain, given by
\begin{equation}\label{eq:initial0}
    \phi_0(x, y) = \sqrt{x^2 + y^2} - R,
\end{equation}
with $R = 0.8$ the radius of the 0-level set. The velocity field is set to $F = -1$ in the whole domain and we run the simulation until the final time $t_f = 0.3625$ for different resolutions N = 16, 32, 64 and different CFL numbers ranging from 1 to 16. The $L_2$ norm of the error with respect to the analytical solution is computed in the whole domain. Results in Fig.~\ref{fig:CFL} show a second-order accuracy for any given CFL number for the retracting circle case.

\subsection{Reinitialisation step}
\label{subsec:Reinitialisation}

\begin{figure*}[ht!]
        \centering
        \includegraphics[width=1\textwidth]{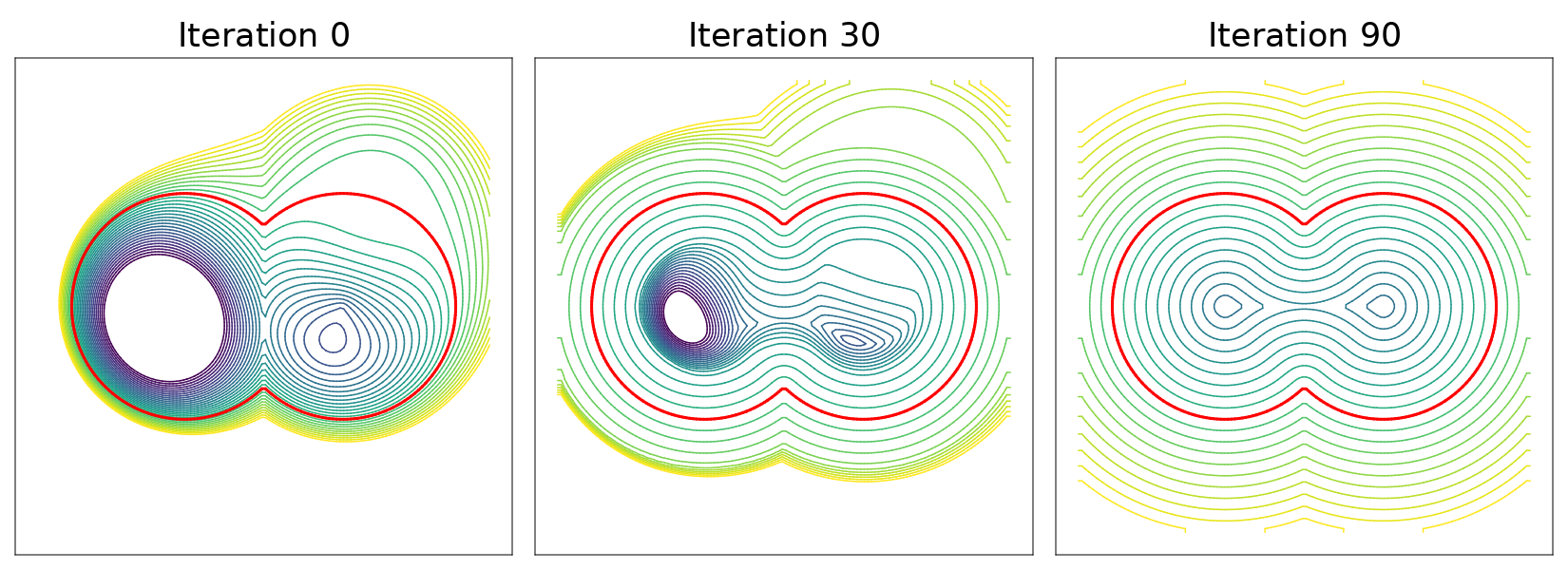}
        \caption{The -2 to 1 level sets of the function $\phi$ are shown with a $0.1$ step. The red curve represents the 0-level set. After 90 iterations the level set function is a true signed distance function.}
        \label{fig:reinit}
\end{figure*}

Depending on the case, usually after one or more iterations of the time advancement scheme, the level set function will cease to be an exact signed distance function, necessitating a reinitialisation step to enforce the level set function $\phi$ to remain an exact distance function from the front $\Gamma \left ( t \right )$. Given a function $\phi_0$, which is not a signed distance function, it can be evolved into a function $\phi$ by solving the Eikonal equation
\begin{equation}\label{eq:eikonal}
\left\{\begin{array}{rr}
\dfrac{\partial \phi}{\partial \tau}=\operatorname{S}\left(\phi_{0}\right)(1-|\nabla \phi|)&\text{in} \: \Omega \\
\phi(x, \tau) = 0 & \text{on} \: \Gamma \left ( t \right )
\end{array}
\right.
\end{equation} 
where $\tau$ denotes a pseudo-time, $S(\phi_ 0)$ denotes a sign function, and $\phi$ a signed distance function (once steady-state is reached). There exist many numerical methods for solving the Eikonal equation and, here we have adopted that of Min \textit{et al.}~\cite{min_reinitializing_2010}, also recently used in the work of Limare \textit{et al.}~\cite{Limare2022}. This approach relies on a second order ENO spatial discretization with sub-cell resolution near the interface. The one-sided ENO finite difference (in the $x$ direction only) yields
\begin{equation}\label{eq:enoshifted}
\begin{aligned}
&D_{x}^{+} \phi_{i j}=\frac{\phi_{i+1, j}-\phi_{i j}}{h}-\frac{h}{2} \operatorname{minmod}\left(D_{x x} \phi_{i j}, D_{x x} \phi_{i+1, j}\right), \\
&D_{x}^{-} \phi_{i j}=\frac{\phi_{i, j}-\phi_{i-1, j}}{h}+\frac{h}{2} \operatorname{minmod}\left(D_{x x} \phi_{i j}, D_{x x} \phi_{i-1, j}\right)
\end{aligned}
\end{equation}
where $D_{x x} \phi_{i j}=\left(\phi_{i-1, j}-2 \phi_{i j}+\phi_{i+1, j}\right) /h^2$ is the second order derivative of $\phi_{ij}$ and the ``minmod'' limiter is zero when both arguments have opposite signs, and takes the argument of smallest absolute value otherwise. The numerical Hamiltonian $|\phi|$ is computed as follows
\begin{equation}\label{eq:hamiltonian}
|\nabla \phi|_{i j} \simeq H_{G}\left(D_{x}^{+} \phi_{i j}, D_{x}^{-} \phi_{i j}, D_{y}^{+} \phi_{i j}, D_{y}^{-} \phi_{i j}\right)
\end{equation}
where function $H_g$ is given by
\begin{equation}
H_{G}(a, b, c, d)= \begin{cases}\sqrt{\max \left(\left(a^{-}\right)^{2},\left(b^{+}\right)^{2}\right)+\max \left(\left(c^{-}\right)^{2},\left(d^{+}\right)^{2}\right)} \quad \text{when} \quad \operatorname{S}\left(\phi^{0}\right) \geqslant 0, \\ \sqrt{\max \left(\left(a^{+}\right)^{2},\left(b^{-}\right)^{2}\right)+\max \left(\left(c^{+}\right)^{2},\left(d^{-}\right)^{2}\right)} \quad \text{when} \quad \operatorname{S}\left(\phi^{0}\right)<0.\end{cases}
\end{equation}
Near the interface, the finite differences need to be modified in order to impose $\phi = 0$ where $\phi_0 = 0$. To this end, a quadratic ENO polynomial interpolation gives 
\begin{equation}\label{eq:modified_eno}
    D_{x}^{+} \phi_{i j}=\frac{0-\phi_{i j}}{h^{+}}-\frac{ h^{+}}{2} \operatorname{minmod}\left(D_{x x} \phi_{i j}, D_{x x} \phi_{i+1, j}\right)
\end{equation}
where 
\begin{equation}\label{eq:hplus}
\Delta h^{+}= \begin{cases}h\left(\frac{1}{2}+\frac{\phi_{i j}^{0}-\phi_{i+1, j}^{0}-\operatorname{S}\left(\phi_{i j}^{0}-\phi_{i+1, j}^{0}\right) \sqrt{D}}{\phi_{x x}^{0}}\right) \quad \text{where} \quad \left|\phi_{x x}^{0}\right|>\epsilon, \\ h \frac{\phi_{i j}^{0}}{\phi_{i j}^{0}-\phi_{i+1, j}^{0}} \quad \text{elsewhere} \end{cases}
\end{equation}
with $\phi_{x x}^{0}=\operatorname{minmod}\left(\phi_{i-1 j}^{0}-2 \phi_{i j}^{0}+\phi_{i+1, j}^{0}, \phi_{i j}^{0}-2 \phi_{i-1, j}^{0}+\phi_{i-2 j}^{0}\right)$ and $D=\left(\phi_{x x}^{0} / 2-\phi_{i j}^{0}-\phi_{i-1, j}^{0}\right)^{2}-4 \phi_{i j}^{0} \phi_{i-1, j}^{0}$. The negative one-sided ENO difference $D_{x}^{-}$ is obtained similarly. A forward Euler scheme is then used for time advancement. This Hamiltonian extraction will also be used for check-pointing in the adjoint problem (Sec.~\ref{sec:Optimality}) when needed. We validate the method by initializing a perturbed solution, similar to the test case found in ~\cite{min_reinitializing_2010}, in a $4 \times 4$ domain
\begin{equation}
\phi_0(x,y) = ((x - 1)^2 + (y - 1)^2 + 0.1) \times (\sqrt{(x \pm a)^2 + y^2}-R)
\end{equation}
where $R = 1$ and $a = 0.7$. Fig.~\ref{fig:reinit} shows the initial perturbed level set function in a $128 \times 128$ grid converging towards a signed distance function with increasing number of iterations in time.
\subsection{Dead and fresh cells}
\label{subsec:Deadfresh}
\begin{figure*}[ht!]
\begin{center}
\begin{tikzpicture}[scale=2.3]

\draw[fill = blue!10] (0,0) rectangle (3,3);
\draw[fill = brown!10] (1.7, 3) -- (1.3, 0) -- (0,0) -- (0,3);

\foreach \ii in {0, 1, 2, 3}
    	{\draw [very thick] (0, \ii) to (3,\ii);}
\foreach \ii in {0, 1, 2, 3}
    	{\draw [very thick] (\ii, 0) to (\ii, 3);}

\draw [very thick, red] (1.7, 3) to (1.3, 0);

\foreach \ii in {0.5, 1.5, 2.5}
	{\node at (0.5,\ii) [circle,fill,inner sep=2.5pt]{};
	\node at (1.5,\ii) [rectangle,fill,inner sep=3.5pt]{};
	\node at (2.5,\ii) [cross, fill, inner sep=3.5pt]{};}

\draw [-{Stealth[length=3mm, width=2mm]}, very thick, red] (1.47, 1.3) to (1.9, 1.21);
\node at (1.8, 1.31) [above] {\Large{\textcolor{red}{v}}};

\draw (0.5, 2.8) node[] {\large{Solid}};
\draw (2.5, 2.8) node[] {\large{Liquid}};

\draw (1.5, 3) node[above, yshift=5.5pt] {\Large{Stencil at time $t^{n}$}};

\node at (0.2,0) [below, yshift=-5.5pt, circle, fill, inner sep=2.5pt]{}; 
\node at (0.25,0) [right, yshift=-8.75pt] { : \footnotesize{full cells}};

\node at (1.1,0) [below, yshift=-5.5pt, rectangle, fill, inner sep=3.5pt]{}; 
\node at (1.15,0) [right, yshift=-9.75pt] { : \footnotesize{partial cells}};

\node at (2.1,0) [below, yshift=-5.5pt, cross, fill, inner sep=3.5pt]{}; 
\node at (2.15,0) [right, yshift=-9.75pt] { : \footnotesize{empty cells}};

\draw [-{Stealth[length=3mm, width=2mm]}, very thick, black] (1.55, 1.75) to (1.8, 1.7);
            
\draw (1.675, 1.725) node[above, yshift=2pt] {$n$};

\begin{scope}[xshift = 3.5cm]
\draw[fill = blue!10] (0,0) rectangle (3,3);
\draw[fill = brown!10] (2.7, 3) -- (2.3, 0) -- (0,0) -- (0,3);

\foreach \ii in {0, 1, 2, 3}
    	{\draw [very thick] (0, \ii) to (3,\ii);}
\foreach \ii in {0, 1, 2, 3}
    	{\draw [very thick] (\ii, 0) to (\ii, 3);}

\draw [very thick, red] (2.7, 3) to (2.3, 0);

\foreach \ii in {0.5, 1.5, 2.5}
	{\node at (2.5,\ii) [rectangle, fill, inner sep=3.5pt]{};}

\foreach \ii in {0.5, 1.5, 2.5}
            	{\FPeval{\result}{clip(0.5 + \ii)}%
            	\node at (0.5,\ii) [circle,fill,inner sep=2.5pt]{};
            	\node at (1.5,\ii) [circle,fill,inner sep=2.5pt]{};
            	\draw (0.5, \ii) node[right, xshift=5pt] {$T^{\result}_B$};
            	\draw (1.5, \ii) node[right, xshift=5pt] {$T^{\result}_A$};}

\draw [very thick, dotted] (0.5, 2.5) to (0.5, 0.5);
\draw [very thick, dotted] (1.5, 2.5) to (1.5, 0.5);

\node at (2.25, 1.7) [rectangle,fill,inner sep=3.5pt, red]{};
\draw [-{Stealth[length=3mm, width=2mm]}, dashed, very thick,black]   (0.5, 2.1) to (2.25, 1.7);

\draw [-{Stealth[length=3mm, width=2mm]}, very thick, black] (2.55, 1.75) to (2.8, 1.7);

\draw (2.675, 1.725) node[above, yshift=2pt] {$n$};

\node at (0.5, 2.1) [rectangle,fill,inner sep=3.5pt, blue]{};
\node at (1.5, 1.87) [rectangle,fill,inner sep=3.5pt, violet]{};

\draw (2.25, 1.7) node[above, yshift=4.5pt, red] {\large{$T_{new}$}};
\draw (0.5, 2.1) node[left, yshift=10.5pt, blue] {\large{$T_B$}};
\draw (1.5, 1.87) node[below, xshift=-15.5pt, violet] {\large{$T_A$}};

\draw (1.5, 3) node[above, yshift=5.5pt] {\Large{Stencil at time $t^{n+1}$}};
\end{scope}

\end{tikzpicture}
\end{center}
    \caption{Example of empty cells becoming partial cells from the point of view of the solid phase. The interface at time $t^{n}$ moves in the normal direction with speed $V$. At time $t^{n+1}$, the newly initialized value $T_{new}$ located at the partial cell centroid is extrapolated from $T_A$ and $T_B$.}
    \label{fig:fresh_dead}
\end{figure*}
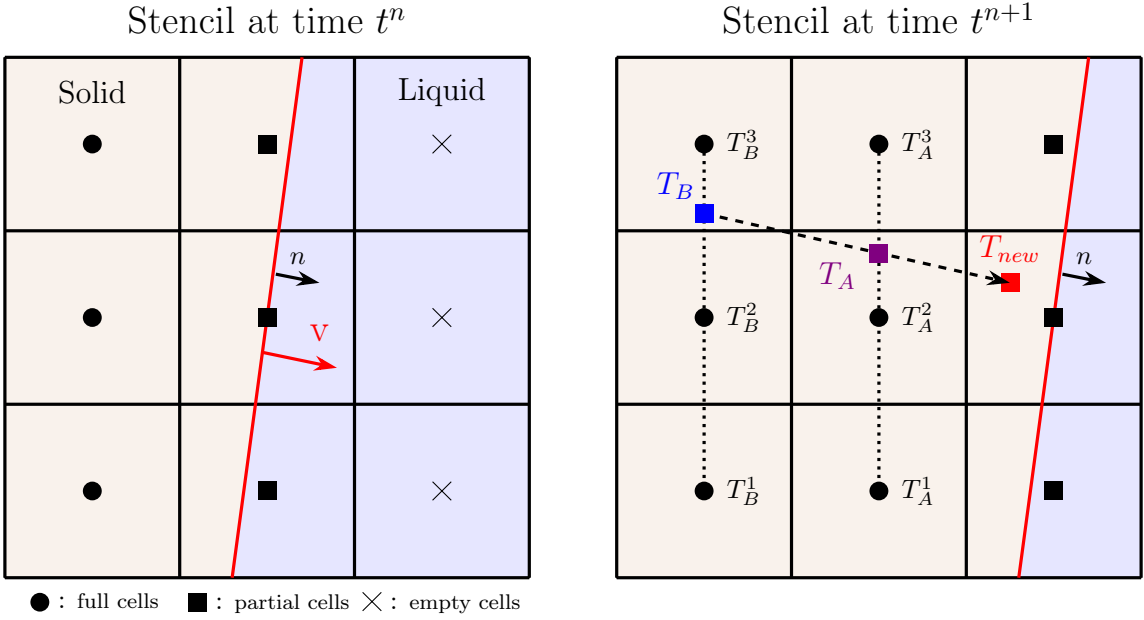

The last step of the method presented in this study is the treatment of fresh and dead cells. As the interface moves through the Cartesian grid, the temperature field needs to be initialized according to the front position. Two cases can be distinguished: (i) a full or partial cell becomes an empty cell, and (ii) an empty cell becomes a partial cell. In the first case, the previous temperature value is simply set to 0. In the second case, the temperature previously non-existing needs to be initialized. The procedure is similar to that of the Stefan condition, Sec.~\ref{subsec:Stefan}: (i) a shifted $3\times 3$ stencil, as shown in Fig.~\ref{fig:fresh_dead}, is chosen, (ii) a line from the interface centroid is cast in the opposite normal direction $-n$, (iii) the crossing points $A$ and $B$ of this line and the vertical (or horizontal depending on the normal orientation) segments of the neighboring 3 points, are identified, (iv) the values $T^*_A$ and $T^*_B$ are interpolated using $T^1_A, T^2_A, T^2_A$ and $T^1_B, T^2_B, T^2_B$ respectively, and (v) the coordinates $\mathbf{o} = (x_\mathrm{new}, y_\mathrm{new})$ of the barycentre of the cell to initialize is located. This last step is done by using the discrete face capacities defined in Sec.~\ref{subsec:Temperature}. An orthonormal coordinate system $\mathcal{R} = (\mathbf{o}, (x^{\prime},n))$ similar to the parabola-fitted curvature found in~\cite{popinet_accurate_2009} is then defined. Finally, the new temperature value, $T_\mathrm{new}$, is linearly extrapolated by solving for $a_0$ and $a_1$ in the resulting system of equations
\begin{equation}\label{eq:system_fresh}
    \left\{\begin{aligned}
    T^*_A = a_0 x^{\prime}_A + a_1 \\
    T^*_B = a_0 x^{\prime}_B + a_1
    \end{aligned}
    \right.
\end{equation}
where $a_1 = T_\mathrm{new}$.

\section{Validation and assessment}
\label{sec:Validation}
In this section, the numerical scheme previously described is validated and the effect of grid refinement, as well as, various physical parameters of the system, such as surface tension at the interface, on the resulting crystal growth are demonstrated and discussed.   
\subsection{Frank's sphere}

The growth of an ice sphere surrounded by an under-cooled liquid was initially studied by Frank~\cite{Frank1950}, and the initial sphere radius was shown to evolve in a self-similar manner as the square-root of time. This problem can therefore be used to validate the accuracy of the numerical scheme implemented here. 

\begin{figure*}
\begin{center}
	\includegraphics[width=1\textwidth]{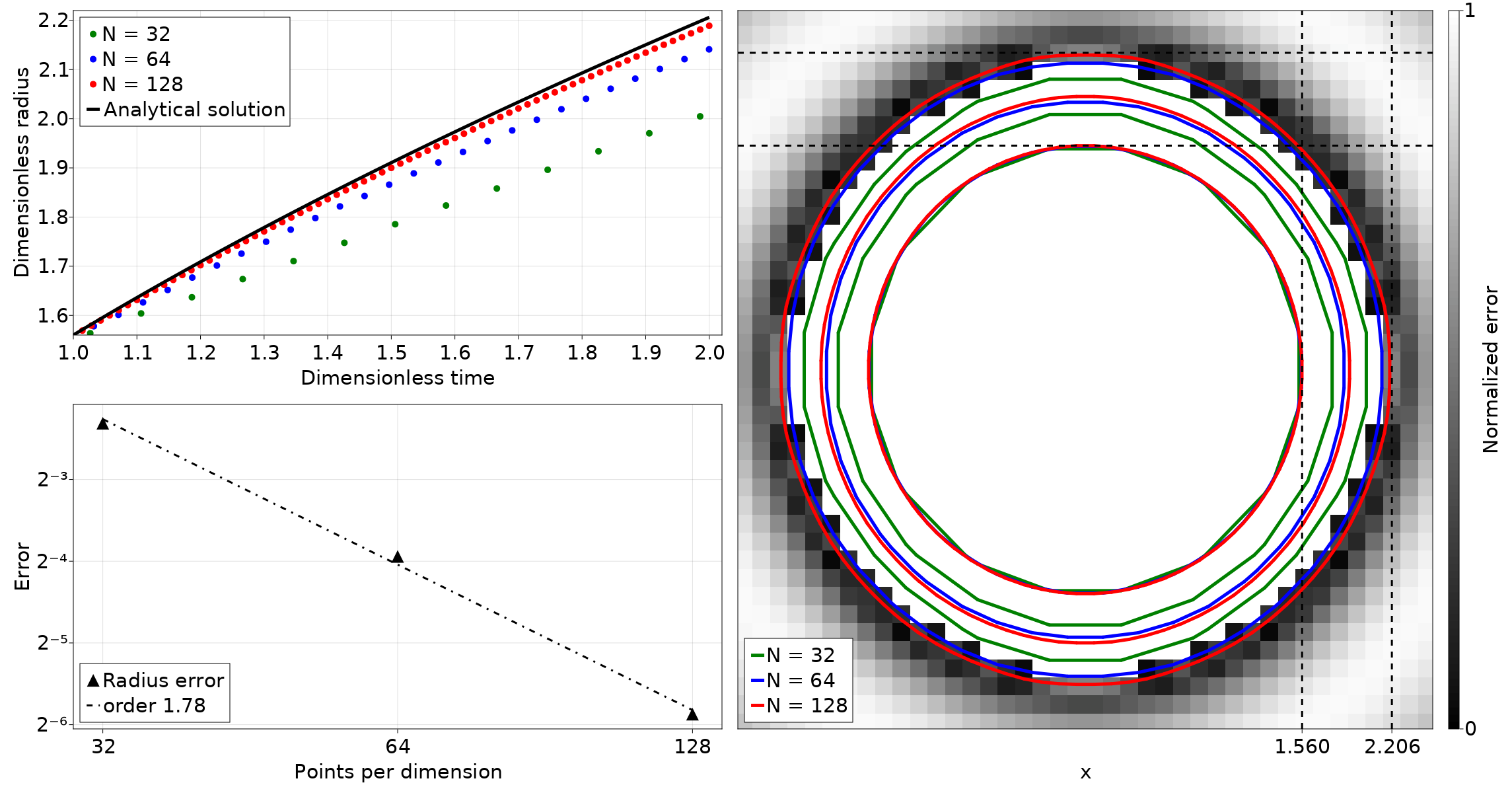}
    \caption{Convergence of the growing Frank's sphere for different resolutions N = 32, 64, 128. The top left figure shows the convergence of the radius towards the analytical solution as a function of time. The bottom left figure represents the error of the final radius as a function of $N$. The right figure shows the position of the interface at times $t = 0, 0.5, 1$ for the different grids. The dotted lines correspond to the initial radius $R_0 = 1.56$ and the analytical radius at final time $R(t = 1) = 2.206$.}
\label{fig:Frank}
\end{center}
\end{figure*}

In this configuration, the temperature field is given by
\begin{equation}\label{eq:similarity_frank}
T(r, t) = T(s) = \left\{
\begin{array}{lr} 
T_\infty \left( 1 - \dfrac{F(s)}{F(S)} \right), & s > S \\
0, & s < S
\end{array}\right.
\end{equation}
where $r = \sqrt{x^2 + y^2}$, $s = r/t^{1/2}$, $T_\infty$ is a given under-cooling temperature, and $F(s)$ denotes the similarity solution of the heat equation $F(s) = \operatorname{E}_1 (1/4 \: s^2)$. Numerical errors may lead to an unwanted alteration of the initial shape due to the Mullins-Sekerka instability~\cite{mullins_stability_1964}, where a perturbed solution can lead to unstable dendritic growth in the case of zero melting temperature ($T_M = 0$). We therefore test the robustness of our method using the initial parameters recommended in Almgren \textit{et al.}~\cite{ALMGREN1993}. The initial level set function is set to a circle of radius $R_0 = 1.56$ in a $8 \times  8$ domain surrounded by an initial temperature field (Eq.~\ref{eq:similarity_frank}) with $T_\infty = -0.5$. The initial time is set to 1 and the simulation is advanced until a final time $t_f = 2$. The results of this study are summarized in Fig.\ref{fig:Frank}. The intrinsic regularization of our method allows the level set function to retain its initial circular shape, avoiding spurious interface oscillations. The following test cases will focus instead on the dendritic growth of an initial crystal shape in a uniform under-cooled temperature field.
\subsection{Dendritic growth}

\begin{figure*}[ht!]
        \centering
        
        \includegraphics[width=1.0\textwidth]{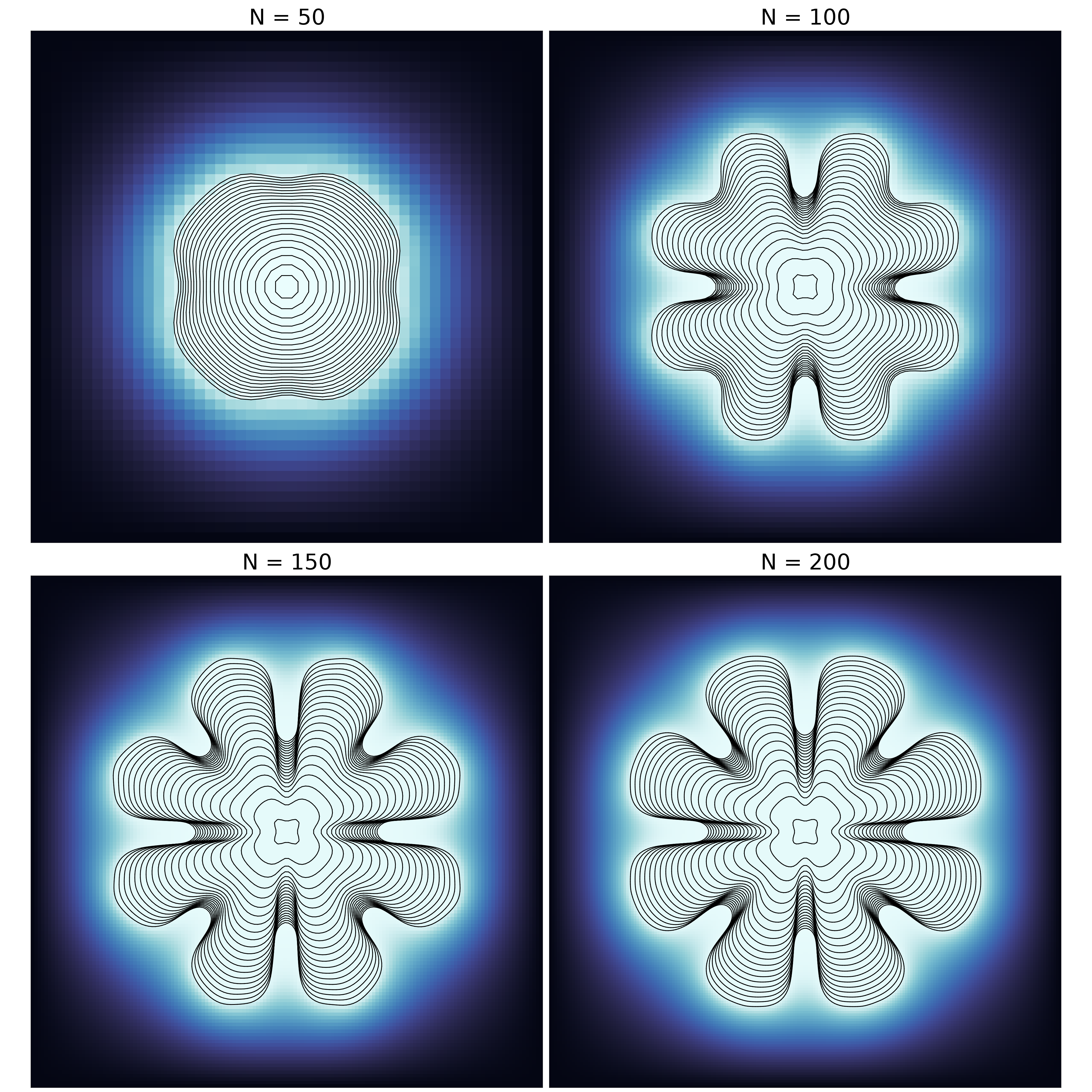}
        \caption{Effect of grid refinement on the dendrite growing tips. The initial condition is a solid crystal at temperature $T = 0$ surrounded by an under-cooling temperature $T_\infty = -0.5$. The color map represents the final temperature field. The final time is $t_f = 0.5$ and the interface is plotted with a time step of $0.025$.}
        \label{fig:crystal_grid}
\end{figure*}

Crystal growth is an unstable phenomenon that occurs spontaneously in nature. Its appearance is the result of a competition between the natural growth due to the Stefan condition (Eq.~\ref{eq:stefan}), which tends to stretch the tips, and the surface tension effect present in the Gibbs-Thomson relation (Eq.~\ref{eq:gibbs-thomson}), which tends to restore the flatness of the interface (see~\cite{Langer1980} for further details). Numerical reproduction of such patterns is a challenge, since the dendritic growth depends on the smallest resolved length scale. In order to validate our method, a case similar to that of Chen \textit{et al.}~\cite{chen_simple_1997} is considered, where an initial solid crystal is surrounded by an under-cooled liquid. The level set function is initialized in a $2 \times 2$ domain as
\begin{equation}\label{eq:crystal}
    \phi_0 (x,y) = \sqrt{x^2 + y^2} \times (0.1 + 0.02\operatorname{cos}(4 \alpha) - 0.01),
\end{equation}
where $\alpha$ is the angle of the interface with respect to the $x$ axis, and the under-cooled temperature is $T_\infty = - 0.5$. 

At the first stage, the effect of the grid resolution on the solution (Fig.~\ref{fig:crystal_grid}) is assessed. Surface tension coefficient is fixed at $\epsilon_\kappa = 0.0004$, and velocity coefficient is set to $\epsilon_V = 0$. The simulations are advanced with insulated boundary conditions until a final time $t_f = 0.5$ for different grid resolutions N = 50, 100, 150, 200. Fig.~\ref{fig:crystal_grid} shows that as the grid size decreases, the tip-splitting appears earlier and the length of the dendrites increases. The final shape converges towards an 8-fold symmetric crystal shape. 

\begin{figure*}[ht!]
        \centering
        
        \includegraphics[width=1.0\textwidth]{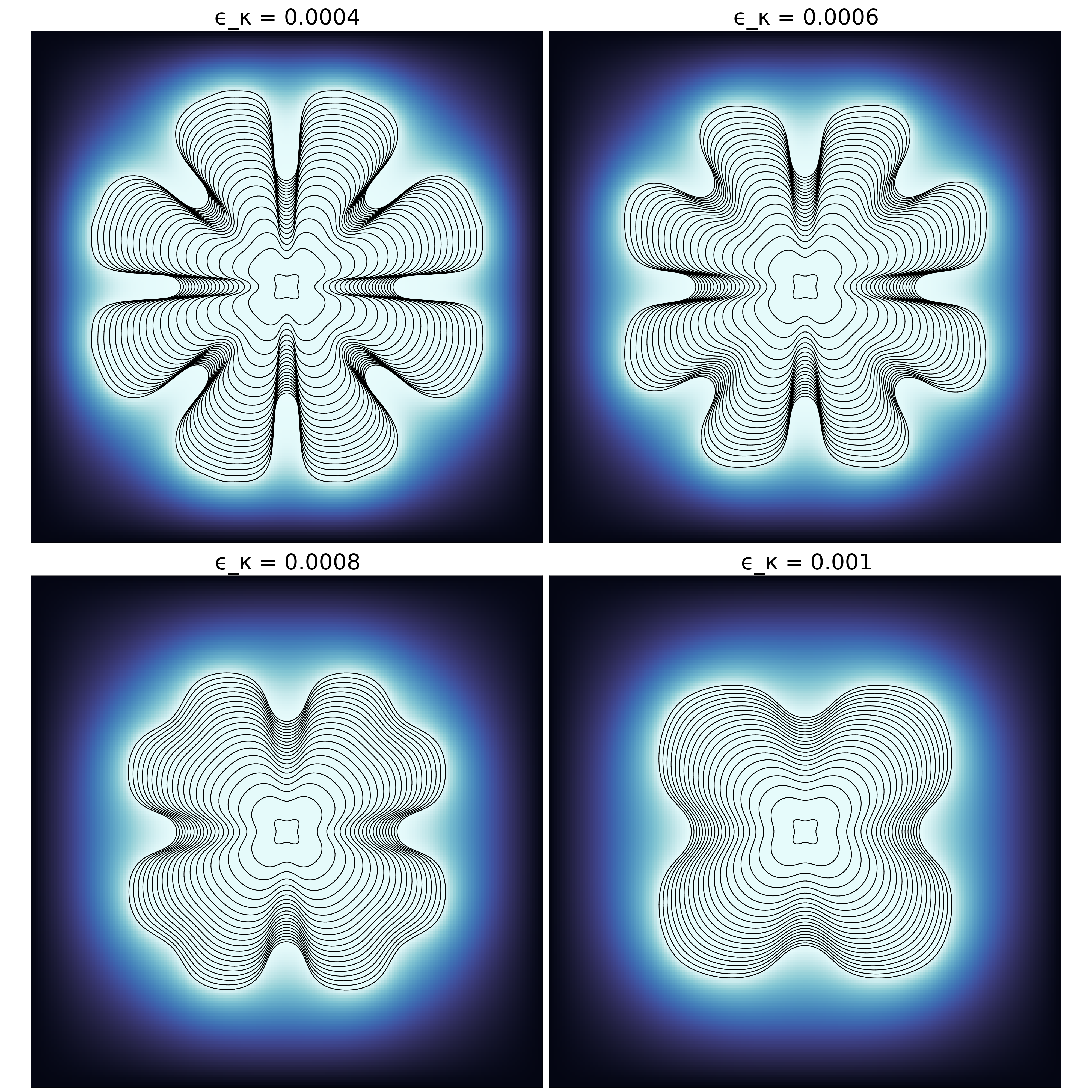}
        
        \caption{Effect of the surface tension coefficient on the dendrite growing tips. The initial condition is a solid crystal at temperature $T = 0$ surrounded by an under-cooling temperature $T_\infty = -0.5$. The color map represents the final temperature field. The final time is $t_f = 0.6$ and the interface is plotted with a time step of $0.03$.}
        \label{fig:crystal_surfacetension}
\end{figure*}

In the second case, the surface tension coefficient is varied from $0.0004$ to $0.001$ for a fixed $N = 200$. Fig.~\ref{fig:crystal_surfacetension} clearly demonstrates the stabilizing effect of the Gibbs-Thomson relation. The tip-splitting disappears as the surface tension increases. This behaviour is explained by a stronger regularization due to higher surface tension coefficient, reducing the growth rate of the instability, and is in agreement with the observations in similar regimes~\cite{chen_simple_1997}.

\subsection{Anisotropy effects}

\begin{figure*}[ht!]
    \centering
    \subfigure[$\alpha_0 = \pi/2$]
    {\includegraphics[width=0.45\textwidth]{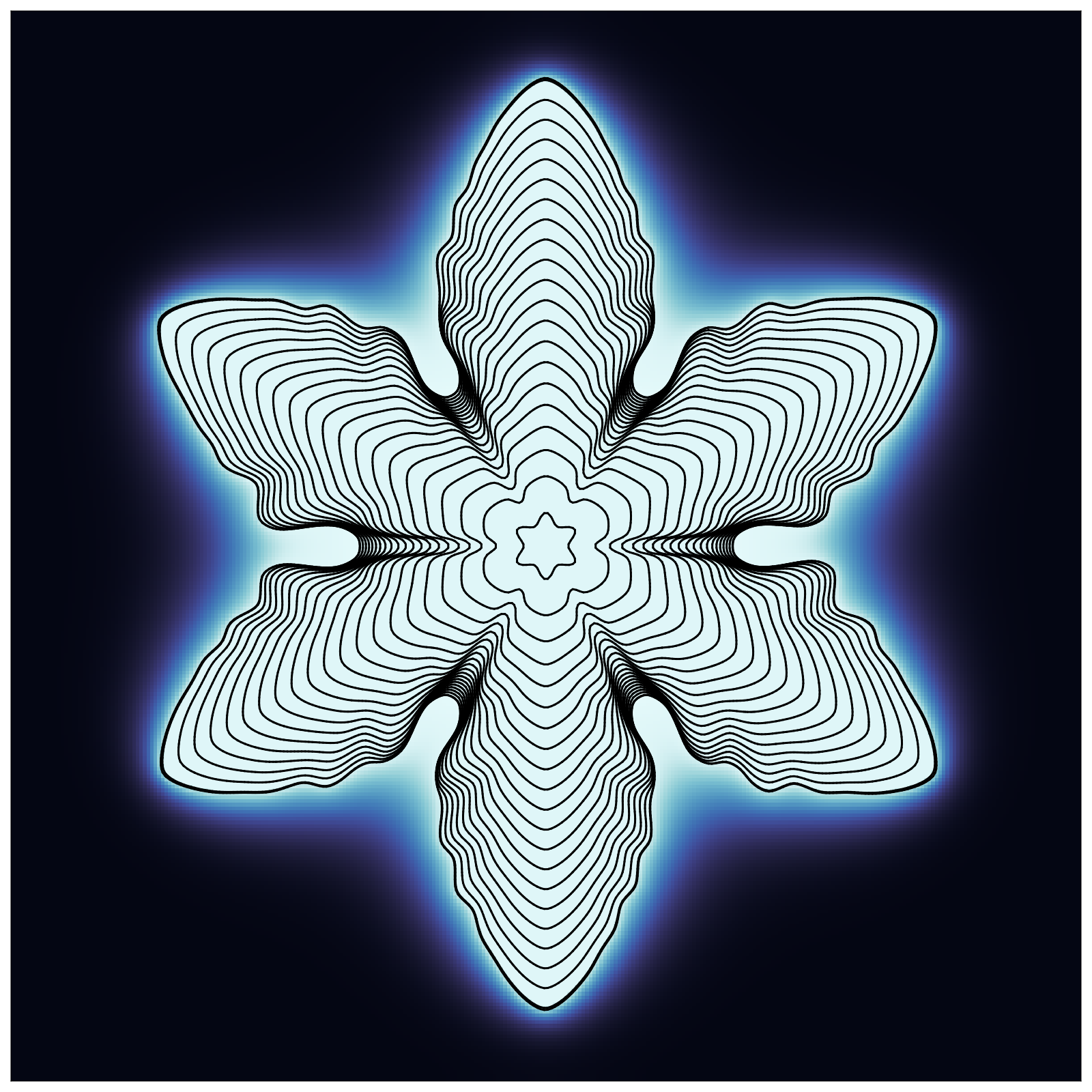}
    \label{fig:anisotropy_pi2}
    }
    \subfigure[$\alpha_0 = \pi/4$]
    {\includegraphics[width=0.45\textwidth]{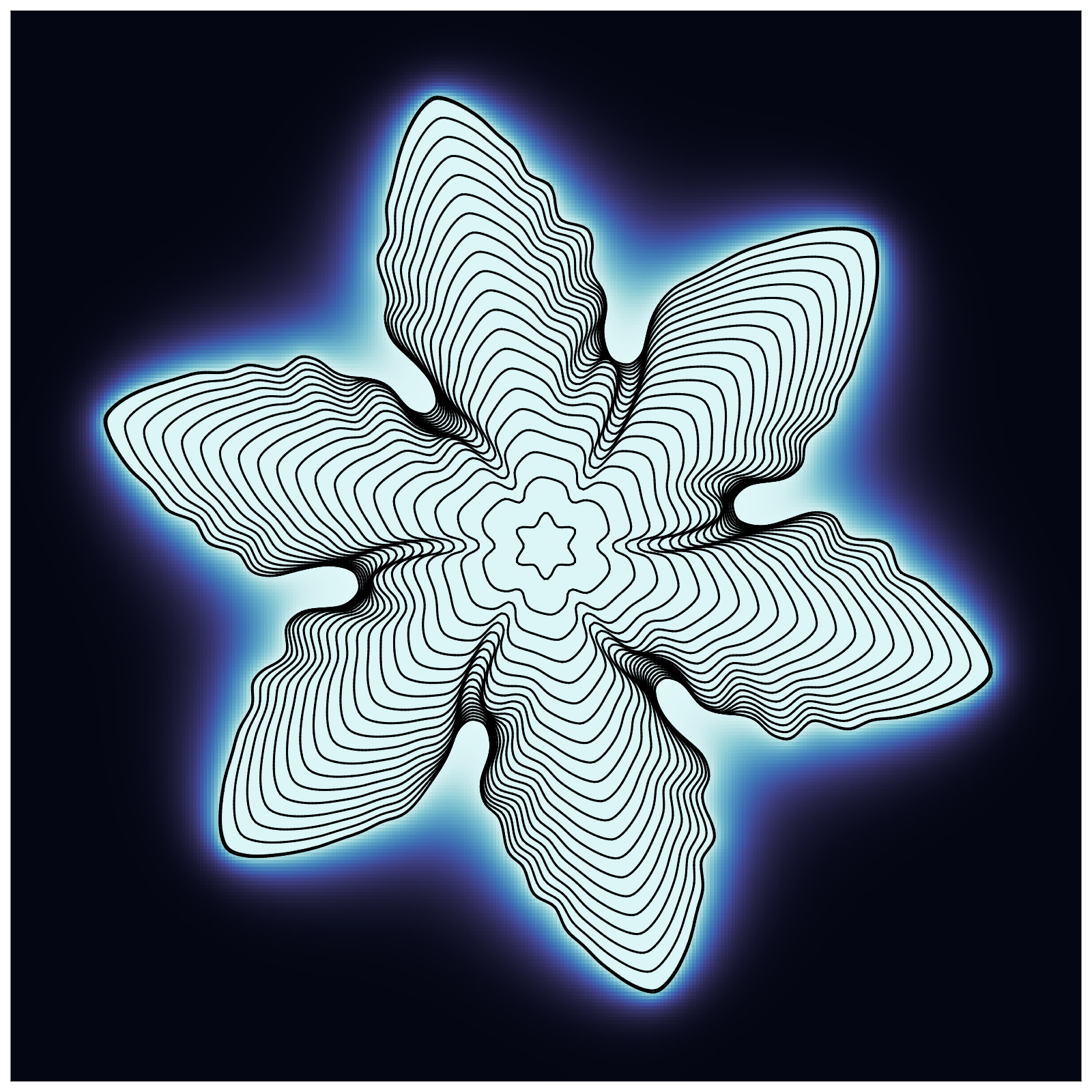}
    \label{fig:anisotropy_pi4}
    }
  \caption{Anisotropy effects on the crystal growth for different prescribed angles $\alpha_0 = \pi/2, \pi/4$. The initial condition is a six-fold solid at temperature $T = 0$ surrounded by an under-cooling temperature $T_\infty = -0.8$. The color map represent the final temperature field. The final time is $t_f = 0.09$ and the interface is plotted with a time step of $0.0045$.}
  \label{fig:anisotropy}
\end{figure*}

In order to accurately reproduce the crystal shapes produced by dendritic growth, we introduce a variable surface tension coefficient, similar to ~\cite{Tan2006} and ~\cite{Limare2022}, in the Gibbs-Thomson relation (Eq.~\ref{eq:gibbs-thomson}) to account for anisotropy effects,
\begin{equation}\label{eq:anisotropy}
    \Bar{\epsilon}_{\kappa}(\alpha) = \epsilon_{\kappa} \left(1+ A\left[\frac{8}{3} \sin ^{4}\left(\frac{1}{2} M \left(\alpha-\alpha_0\right)\right)-1\right]\right),
\end{equation}
where $\epsilon_\kappa$ is the surface tension coefficient, $A$ represents the weight of the anisotropy effect, $M$ the mode number, $\alpha$ the angle of the interface with respect to the x-axis and $\alpha_0$ the prescribed angle of symmetry on which the dendrites will grow. To validate these effects, we initialize a six-fold crystal in an under-cooled liquid with $T_\infty = -0.8$ and a fixed velocity coefficient $\epsilon_V = 0$. The anisotropic weight is fixed at $A = 0.4$ and mode number $M = 6$ and the simulations are performed for two different prescribed angles $\alpha_0 = \pi/2, \: \pi/4$ until a final time $t_f = 0.09$ in a $2\times2$ domain with $N = 300$. Fig.~\ref{fig:anisotropy} shows the crystals growing in the direction of the prescribed angle as well as secondary dendrites appearing from the main branches. When $\alpha_0$ is different from the initial crystal shape (Fig.~\ref{fig:anisotropy_pi4}), a rotation of the primary dendrites towards that prescribed direction can be observed. 


%
\section{Optimisation algorithm}
\label{sec:Optimality}

In this section, the optimization problem corresponding to the forward problem (\ref{FP}) presented in Sec.~\ref{sec:Continuous} is presented, together with the respective cost functional and the corresponding adjoint problem, where the control variable is a Neumann boundary condition that acts on the domain boundary. The resulting optimisation algorithm, used throughout the various cases considered in this study, is also briefly described. The adjoint heat equation and the transport theorem used for the derivation are presented in detail in \ref{app:adjoint_derivation}. For the adjoint level set derivation, we refer the reader to Bernauer \textit{et al.}~\cite{bernauer_optimal_2011}.

We first define $T_{t_{f}}$ and $\phi_{t_{f}}$ as the desired temperature field and level set function respectively. The following tracking-type cost functional provides a mathematical description of the control goals stated above:
\begin{equation}\label{eq:cost_functional}
\mathcal{J} (T, \phi, u)=
\frac{\beta_{1}}{2} \int_{\Omega}\left|T({t_{f}})-T_{t_{f}}\right|^{2} +\frac{\beta_{2}}{2} \int_{\Omega}\left|\phi({t_{f}})-\phi_{t_{f}}\right|^{2}  + \frac{\beta_{3}}{2} \int_{\Gamma} dS +\frac{\beta_{4}}{2} \int_{0}^{t_{f}} \int_{\partial \Omega}|u|^{2}. 
\end{equation}
The first two terms monitor the final temperature distribution and interface position and are used to initialize the adjoint fields. The third term is used in cases where the instabilities induce an increase in interface length while the final term models and penalises the control cost. The constants $\beta_1$ to $\beta_4$ act as weights for each term. We assume that each choice of the control $u$ leads to unique states $T(u)$ and $\phi(u)$. Therefore, the minimization problem (\ref{MP}) reads,
\begin{equation}
\label{MP} \tag{MP}
\begin{array}{c}
\operatorname{min} _{u} \mathcal{J} (T(u), \phi(u), u) \\
\text{subject to (\ref{FP})}
\end{array}
\end{equation}
Let $\theta$ be the adjoint temperature and $\psi$ the adjoint level set function. To compute the gradient of $\mathcal{J}(u)$, we introduce the Lagrange functional $\mathcal{L}$ (we omit $dx$, $dt$ and $ds$ for the sake of brevity),
\begin{equation}\label{eq:Lagrange}
\arraycolsep=2pt\def\arraystretch{2.2}
\begin{array}{l}
\mathcal{L}\left(T, \phi, u, \theta, \theta_D, \theta_I, \psi\right)=\mathcal{J} (T, \phi, u) \\
- \displaystyle \int_{0}^{t_{f}} \int_{\Omega_1} \left( \dfrac{\partial T_{1}}{\partial t} - \Delta T_{1} \right) \theta_1 - \int_{0}^{t_{f}} \int_{\Omega_2} \left( \dfrac{\partial T_{2}}{\partial t} - \Delta T_{2} \right) \theta_2 \\
- \displaystyle \int_{0}^{t_{f}} \int_{\partial \Omega} \left( T - u \right) \theta_D \\
- \displaystyle \int_{0}^{t_{f}} \int_\Gamma \left( T - \left(T_M -\epsilon_V V -\epsilon_\kappa \kappa\right) \right) \theta_I \\
- \displaystyle \int_{0}^{t_{f}} \int_\Gamma \left(\dfrac{\partial \phi}{\partial t} + [\nabla T_i]^1_2 \cdot \nabla \phi  \right) \psi
\end{array}
\end{equation}
The adjoint system is obtained by setting $\mathcal{L}_T(\cdot) = \mathcal{L}_\phi(\cdot) = 0$. The resulting adjoint Stefan problem (\ref{AP}) reads,
\begin{center} 
Find a function $\theta: \Omega \times[t_{f}, 0] \rightarrow \mathbb{R}$ and a function $\psi$ :
$\Omega \times[t_{f}, 0] \rightarrow \mathbb{R}$ such that:
\[\arraycolsep=2pt\def\arraystretch{2.2}
\label{AP} \tag{AP} \left\{
\begin{array}{rcllr} - \dfrac{\partial \theta_{1}}{\partial t} &=& \Delta \theta_{1} & \: \text{in} \: \Omega_1(t) & \text{(AP.a)} \\
- \dfrac{\partial \theta_{2}}{\partial t} &=& \Delta \theta_{2} & \: \text{in} \: \Omega_2(t) & \text{(AP.b)} \\ 
\theta(x,t_f) &=& \beta_1 (T({t_{f}})-T_{t_{f}}) & \: \text{in}  \: \Omega & \text{(AP.c)} \\
\dfrac{\partial \theta(x,t)}{\partial n} &=& 0 & \: \text{on} \: \partial \Omega & \text{(AP.d)} \\
\theta(x,t) &=& \psi |\nabla \phi| & \: \text{on} \: \Gamma(t) & \text{(AP.e)} \\
\dfrac{\partial \psi}{\partial t} + \operatorname{div} (\psi \vec{V}) &=& \dfrac{1}{|\nabla \phi |} \dfrac{\partial T}{\partial n} [\nabla \theta_i]^1_2 \cdot n & \: \text{on} \: \Gamma(t) & \text{(AP.f)} \\
\psi(x,t_f) &=& -\dfrac{\beta_2}{2 } \left( \dfrac{\partial }{\partial n}|\phi_{t_{f}}|^2 + \kappa \left(|\phi_{t_{f}}|^2 + \dfrac{\beta_3}{\beta_2} \right) \right) & \: \text{in} \: \Omega & \text{(AP.g)} \\
0 &=& \beta_4 u + \theta & \: \text{on} \: \partial \Omega & \text{(AP.h)}
\end{array}
\right.
\]
\end{center}
Equations AP.a and AP.b correspond to the heat equations with reverse time direction while Equation AP.c is used to initialize the adjoint temperature field in both phases. In Equation AP.e, the in-homogeneous Neumann boundary condition in $\partial \Omega$ existing in the forward problem (\ref{FP}) is mapped to an homogeneous alternative. The Dirichlet boundary condition at the interface is defined in Equation AP.f where the adjoint temperature is equal to the adjoint level set value $\psi$ augmented by the Hamiltonian of the level set function $|\phi|$. In Equation AP.f, the adjoint level set advection equation is defined. The source term corresponds to the adjoint Stefan condition and takes into account the normal jump in gradient of adjoint temperature across the interface. The adjoint level set function is initialized using Equation (AP.g), and the final Equation (AP.h) represents the optimality condition, where the left-hand side represents the gradient used to update the control variable $u$.\\
Even thought it is known that gradient-based methods exhibit slow convergence, they are the method of choice to prove that the optimal control approach is reasonable due to their straight forward implementation. Therefore, we chose to solve the minimization problem (\ref{MP}) by using the limited memory BFGS (L-BFGS) method, a quasi-Newton method originally described in~\cite{Liu1989}. The main characteristic of this method is that it determines the descent direction by preconditioning the gradient with an approximation of the Hessian matrix. This information is obtained using past approximations (the number of approximations is determined by the memory length parameter which is set to $m = 10$ in all our optimization test cases) as well as the gradient. As an initial guess for the initial Hessian, we use the scaled identity matrix as described in~\cite{Wright2006}. The following algorithm summarizes the L-BFGS method used in our numerical examples,

\begin{algorithm*}[ht!]
    \SetKwData{Left}{left}
	\SetKwData{This}{this}
	\SetKwData{Up}{up}
	\SetKwInOut{Input}{input}\SetKwInOut{Output}{output}
	\Input{$u_0$, $m = 10$}
	\Output{$u$, $T$, $\phi$, $\theta$, $\psi$}
	
	\BlankLine $k \leftarrow 0$, $l \leftarrow 0$

	\While{not converged}{
	\BlankLine Solve the forward Stefan problem (\ref{FP}) for $T^k$ and $\phi^k$ \\
	\BlankLine Solve the adjoint Stefan problem (\ref{AP}) for $\theta^k$ and $\psi^k$\\
	\BlankLine Compute the gradient: $$\nabla \mathcal{J}^k = \beta_4 u^k + \theta^k$$ \\
	\If{$k \geq 1$}{
	$s^{k-1} = \sigma^{k-1} d^{k-1} \quad g^{k-1} = \nabla \mathcal{J}^k - \nabla \mathcal{J}^{k-1}$ \\
	\uIf{$(s^{k-1})^{T} g^{k-1} \leq 0$}{$l \leftarrow 0$}
	\ElseIf{$(s^{k-1})^{T} g^{k-1} > 0$}{$l \rightarrow l + 1$\\
	\If{$l > m$}{Remove $\{ s^{l-m}, g^{l-m} \}$}
	Add $\{ s^{l-m}, s^{l-m}\}$\\
	}}
	\BlankLine Choose an initial approximation to the inverse of the Hessian $H^{k}_{0}$
	\BlankLine Construct the direction $d^k = -H^k \nabla \mathcal{J}^k$
	\BlankLine Determine $\sigma^k$ using a Line Search algorithm with backtracking where $\sigma^k = \operatorname{argmin}\mathcal{J}(u^k + \sigma^k d^{k})$
	\BlankLine Update $u^{k+1} = u^k + \sigma^k d^k$ 
	\BlankLine $k \rightarrow k + 1$
	}	
	\caption{Optimization procedure using L-BFGS method}
\end{algorithm*}
\BlankLine

\section{Results}
\label{sec:results}

\begin{figure*}[ht!]
    \centering
    \subfigure[Melting circle]
    {\includegraphics[width=0.3\textwidth]{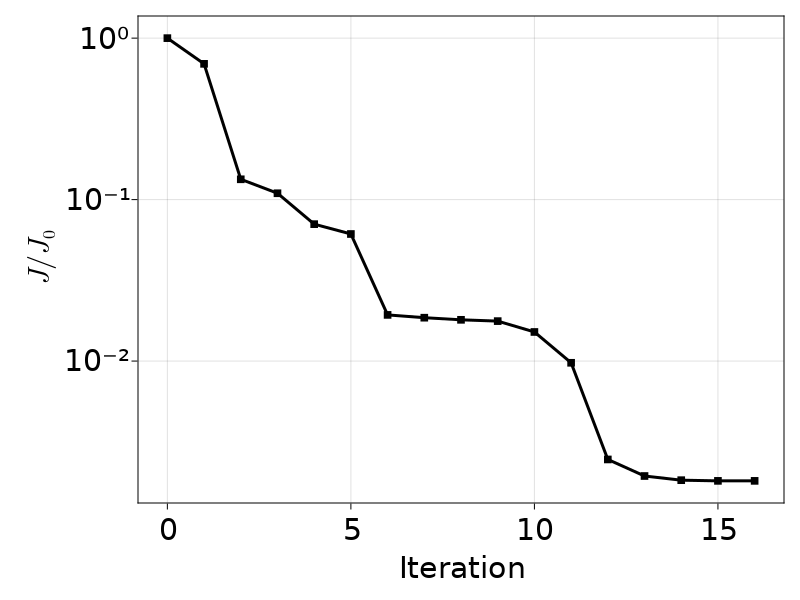}
    \label{fig:cost_meltingcircle}
    }
    \subfigure[Mullins-Sekerka instability]
    {\includegraphics[width=0.3\textwidth]{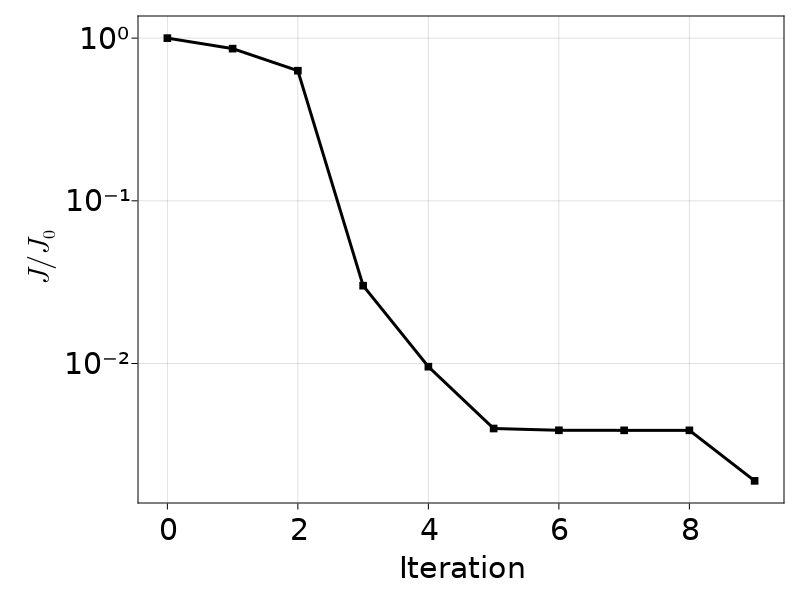}
    \label{fig:cost_mullins}
    }
    \subfigure[Growing crystals]
    {\includegraphics[width=0.3\textwidth]{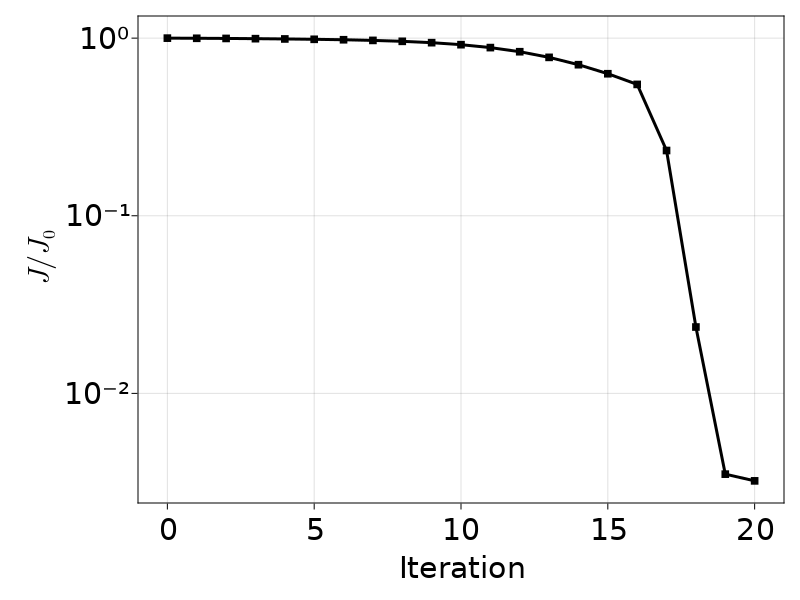}
    \label{fig:cost_crystals}
    }
  \caption{Evolution of the cost functional for the different cases.}
  \label{fig:cost_functionals}
\end{figure*}

A range of optimization cases for the Stefan problem with varying complexity are presented here. The results and performance of the algorithm~\ref{MP} is then analysed for these different setups. In all of the cases the desired temperature field, $T_{t_{f}}$, and level set, $\phi_{t_{f}}$, are computed beforehand and are used to drive the control variable $u$. A summary of the optimization parameters and optimization procedure results for each case is shown in Table~\ref{tab:opt} and the histories of the relative cost functionals $\mathcal{J}/\mathcal{J}^0$ is presented in Fig.~\ref{fig:cost_functionals}.

\paragraph{Implementation:}
The code is written in Julia~\cite{julia2015} a high level scientific programming language. The package \href{https://github.com/flnt/Flower.jl}{\emph{Flower.jl}}, developed by the authors of this study, contains all the methods and test cases presented. We use \emph{Iterative.jl} to solve the linear systems and \emph{Optim.jl}~\cite{Mogensen2018} to build the optimization procedure.
\subsection{Melting circle}

In this case, an initial circle of radius $R = 0.75$ in a $2 \times 2$ domain is considered. The surface tension coefficient is constant and set to $\epsilon_\kappa = 0.002$ while $\epsilon_V = 0$. The non-zero surface tension coefficient is added to further regularize the level set function. The number of points per dimension is $N = 64$ and the final time is set to $t_f = 0.1$. The control $u$ acts on the whole domain boundary and is of Dirichlet type. The adjoint boundary condition, on the other hand, remains a homogeneous Neumann boundary condition. The actuator $u$ is parameterized as 
\begin{equation}\label{eq:basis1}
u = \sum_{p=1}^{2} a_{p} \cos \left(p \pi x\right) 
        + \sum_{p=1}^{2} b_{p} \sin \left(p \pi x\right)
\end{equation}
where $x = [-1, 1]$ corresponds to the bounds of the domain. Through the optimisation process the amplitude of each basis function, $[a_1, a_2, b_1, b_2]$ is prescribed using the gradient equation (AP.h). By opting for a parameterized distribution we ensure the smoothness of the actuation function. Due to the high sensitivity of the cost functional
to the basis considered, with too many parameters creating multiple local minima, the number of parameters are kept at a low enough value to ensure the convexity of the problem while allowing spatial variation of the actuation function. Same strategy will be adopted in the following examples when parametrising the actuation functional. 

The initial guess (ie. the starting point in our cost functional space) is set to $u = 0$ in order not to add a bias regarding the initial descent. Moreover, in this case, the term controlling the interface length $\beta_3$ is set to zero.

\begin{figure*}[ht!]
        \centering
        
        \includegraphics[width=0.8\textwidth]{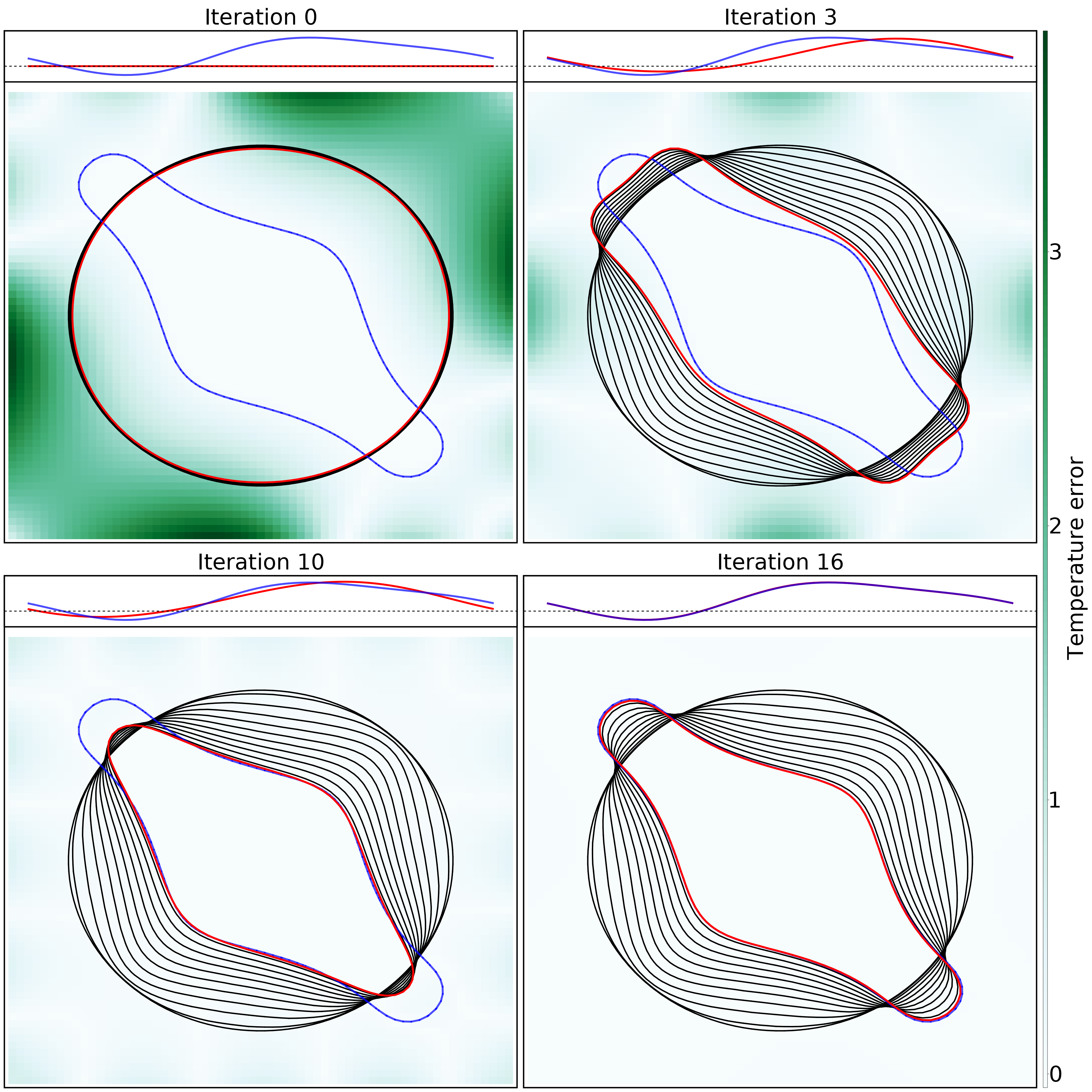}
        \caption{Iterations 0, 3, 10 and 16 of the optimization procedure for the circle test case. The blue curve represents the desired shape and the red one the final position of the interface at a given iteration. The final time is $t_f = 0.1$ and the interface is plotted with a time step of 0.01. The color map corresponds to the error in temperature field. The inset shows the actuator $u$ at a given iteration (red) and the desired one (blue).}
        \label{fig:circle_opt}
\end{figure*}

Fig.~\ref{fig:circle_opt} shows the level set of the forward problem evolving towards the desired shape after a few iterations. The algorithm is able to recover both the retraction and the expansion of the initial circle.

\subsection{Mullins-Sekerka instability}

\begin{figure*}[ht!]
        \centering
        
        \includegraphics[width=0.8\textwidth]{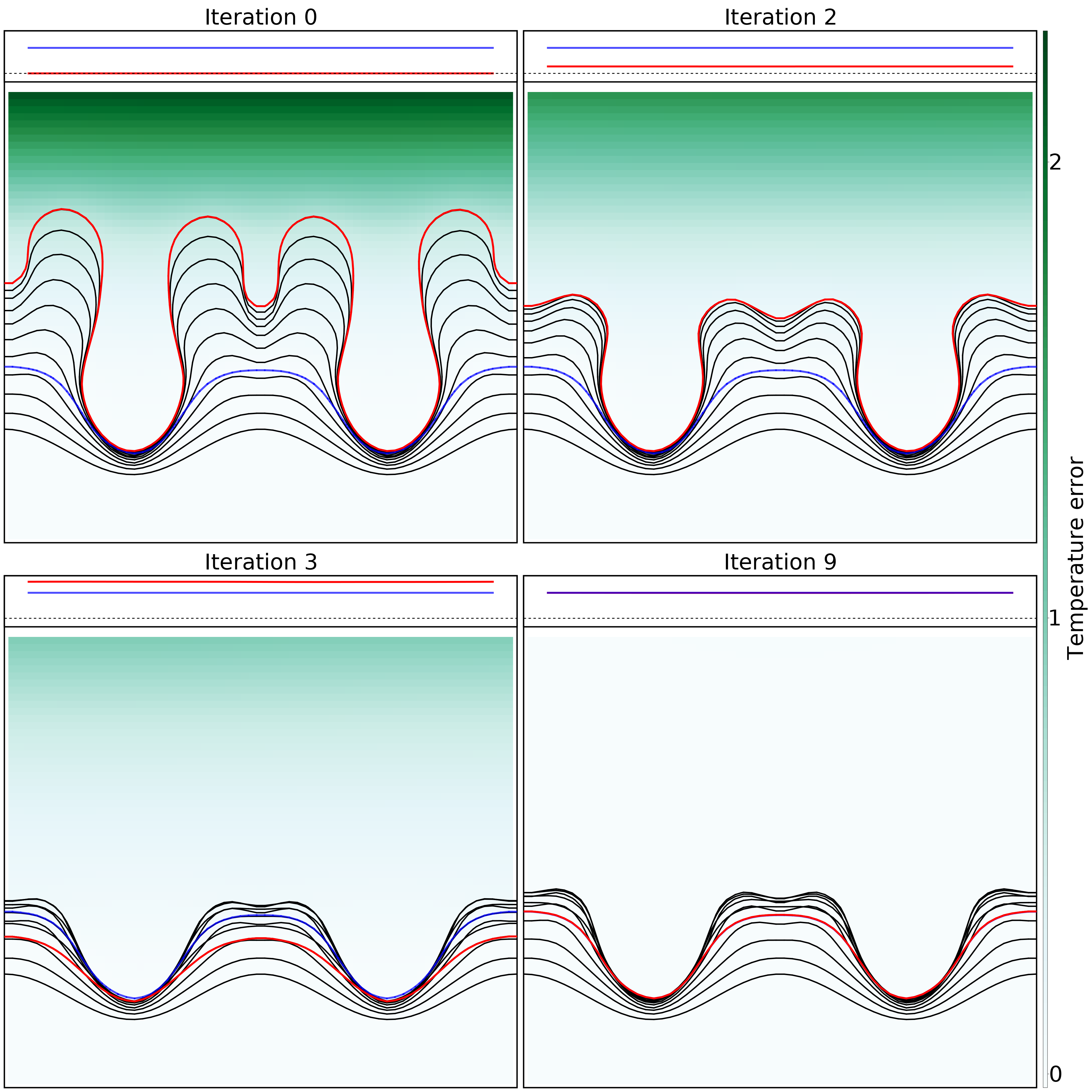}
        \caption{Iterations 0, 2, 3 and 9 of the optimization procedure for the Mullins-Sekerka test case. The blue curve represents the desired shape and the red one the final position of the interface at a given iteration. The final time is $t_f = 0.5$ and the interface is plotted with a time step of 0.05. The color map corresponds to the error in temperature field. The inset shows the actuator $u$ at a given iteration (red) and the desired one (blue).}
        \label{fig:mullins_opt}
\end{figure*}

In this second case, we consider a perturbed initial planar shape in a $2 \times 2$ domain such that the initial level set function is given by,
\begin{equation}\label{eq:initial_mullins}
    \phi_0(x,y) = y + 0.6 + A \operatorname{cos}(2 \pi x)
\end{equation}
and the temperature field by,
\begin{equation}\label{eq:initial_mullins_temp}
T_0(x,y) = \left\{
\begin{array}{lr} 
-1 + \operatorname{e}^{-T_\infty \phi_0(x,y)}, & \phi_0 > 0 \\
0, & \phi_0 < 0
\end{array}\right.
\end{equation}
with the amplitude $A = 0.05$ and the under-cooling temperature $T_\infty = 1.2$.
It is well known that an initial perturbation such as that prescribed in this problem leads to a Mullins-Sekerka type instability~\cite{mullins_stability_1964, chen_simple_1997} characterised by unstable dendritic growth. The surface tension and velocity coefficients are set to zero in this case. The purpose of this optimization test case is reduce the instability by imposing an optimal actuation on the top boundary. The control variable $u$ is of Neumann type and we use $u = 0$ as the initial guess. Both surface tension and velocity coefficients are set to zero and we set the final time $t_f = 0.5$ with $N = 64$. Similar to the previous case the actuation is parametrised as 
\begin{equation}\label{eq:basis2}
u = \sum_{p=1}^{4} a_{p} \cos \left(p \pi x\right) 
        + \sum_{p=1}^{4} b_{p} \sin \left(p \pi x\right)
\end{equation}
In this case, we can add four extra basis function in the actuator parametrization and still recover the global minimum. The extra term in the cost functional (Eq.~\ref{eq:cost_functional}) which controls the length of the interface is also included and $\beta_3 = 0.1$. In Fig.~\ref{fig:mullins_opt}, we can observe how the initial tip splitting is reduced as we go through the optimization procedure. 


\subsection{Anisotropic growing crystals with topology changes}

\begin{figure*}[ht!]
        \centering
        
        \includegraphics[width=1.0\textwidth]{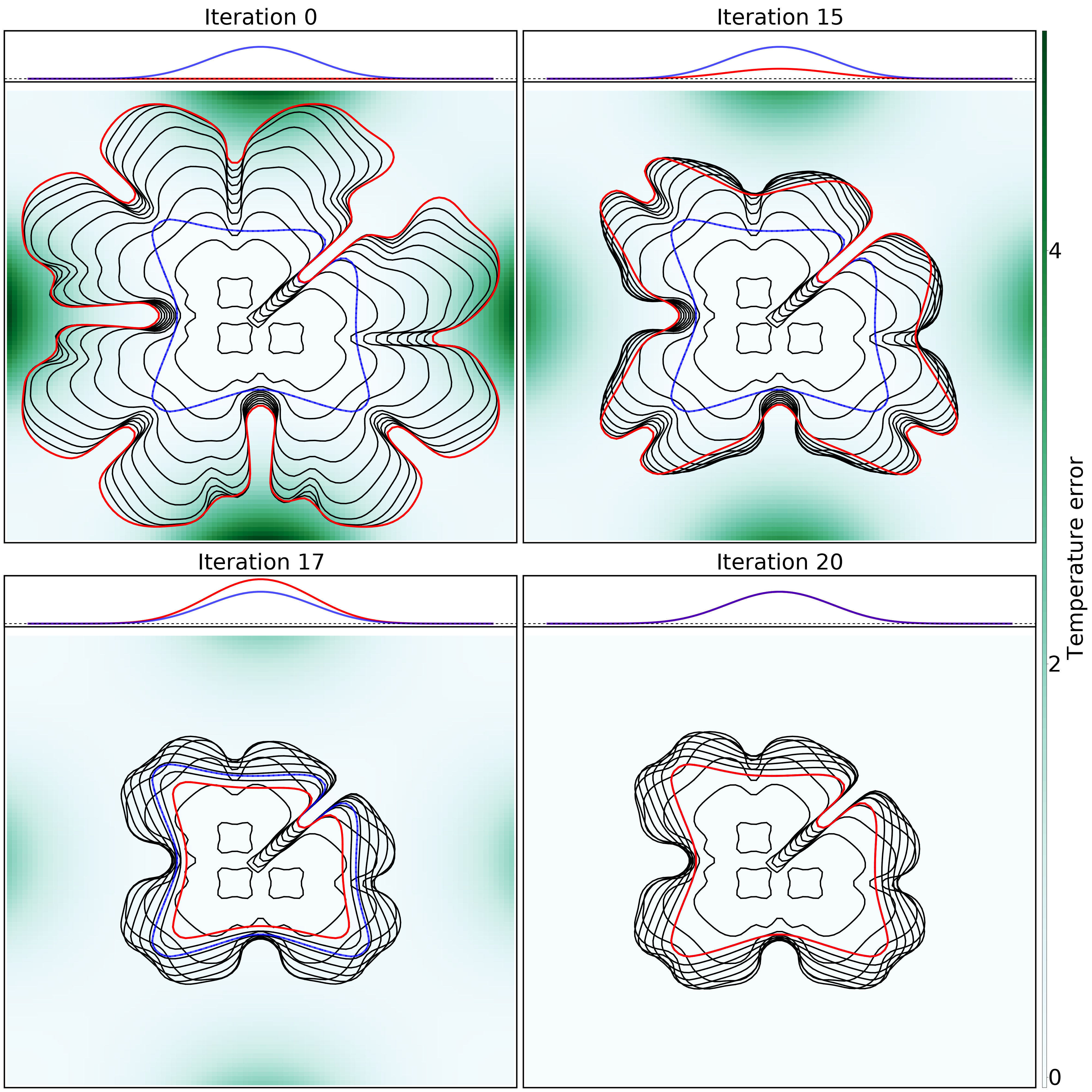}
        \caption{Iterations 0, 15, 17 and 20 of the optimization procedure for the growing crystals test case. The blue curve represents the desired shape and the red one the final position of the interface at a given iteration. The final time is $t_f = 0.45$ and the interface is plotted with a time step of 0.045. The color map corresponds to the error in the temperature field. The inset shows the actuator $u$ at a given iteration (red) and the desired one (blue).}
        \label{fig:crystal_opt}
\end{figure*}

In this last case, we consider three crystals asymmetrically disposed in an under-cooled liquid. The crystals will grow and eventually merge. The objective of this optimization is to drive the final shape towards the desired one by acting on the boundaries of the whole domain, and thereby to suppress the anisotropy effects. The surface tension coeffecient is set to $\epsilon_\kappa = 0.0005$ and the velocity coefficient to $\epsilon_V = 0.002$. We choose a relatively small surface tension coefficient with respect to the velocity one in order to allow for strong dendritic formation and to examine the optimisation algorithm in a challenging case where the topology remains complex. Moreover, we add anisotropy effects by setting $\alpha_0 = \pi/2$ and $M = 4$ in equation\ref{eq:anisotropy}. The heat flux induced through actuation will have to compete with these effects in order to drive the interface towards the desired one. The simulations are run in a $4 \times 4$ domain with $N = 100$ until a final time $t_f = 0.45$. The under-cooling initial temperature is set to $T_\infty = -0.6$. The control variable $u$ is of Neumann type, with an initial guess $u = 0$. The following parametrisation is used, 
\begin{equation}\label{eq:basis3}
u(x,p) = p_1((1 + \operatorname{cos}(\pi/8 x))/2)^4 + p_2((1 + \operatorname{sin}(\pi/8 x))/2)^4
\end{equation}
In this final case, we need to restrict the basis used for the actuator (by fitting only two parameters) in order to descend towards the global minimum. 
Fig.~\ref{fig:crystal_opt} shows the evolution of the surface. It can be seen that initially the crystal is driven towards the domain corners by the anisotropy parameters. Through the optimisation procedure however, the final crystal shape tends towards the desired one. The topology changes are implicitly taken into account by our level set method.

\begin{table*}[hbt]
\setlength{\tabcolsep}{0.5pc}
\newlength{\digitwidth} \settowidth{\digitwidth}{\rm 0}
\catcode`?=\active \def?{\kern\digitwidth}

\begin{tabular*}{\textwidth}{@{}l@{\extracolsep{\fill}}cccccccc}
\hline
                 \multicolumn{1}{c}{Case} 
                 & \multicolumn{4}{c}{Optimization parameters} 
                 & \multicolumn{3}{c}{Results} \\
\cline{2-5} \cline{6-8}
                 & \multicolumn{1}{c}{$\beta_1 $}
                 & \multicolumn{1}{c}{$\beta_2 $} 
                 & \multicolumn{1}{c}{$\beta_3 $} 
                 & \multicolumn{1}{c}{$\beta_4 $}
                 & \multicolumn{1}{c}{$\mathcal{J}$ calls}
                 & \multicolumn{1}{c}{$\nabla \mathcal{J}$ calls} 
                 & \multicolumn{1}{c}{$\mathcal{J}_{final}/\mathcal{J}_{0}$}\\
\hline
Melting circle  & $ 1$ & $10$ & $0$ & $10^{-3}$ & $ 66$ & $16$  & $  1.81 \times 10^{-3}$\\
Mullins-Sekerka & $ 1$ & $10$ &   $0.1$      & $ 10^{-4}$ & $ 54$ &  $  10$ & $ 1.89 \times 10^{-3}$\\
Growing crystals   & $ 1$ & $1$ & $0$ & $10^{-2}$ & $ 25 $ & $ 20 $ & $  3.21 \times 10^{-3}$\\
\hline
\end{tabular*}
\caption{Optimization parameters and final results of the different considered cases. The columns $\mathcal{J}$ and $\nabla \mathcal{J}$ correspond to the number of \ref{FP} and \ref{AP} calls respectively.}
\label{tab:opt}
\end{table*}

\section{Conclusion}
\label{sec:Conclusions}
In this study, we proposed a simple level set method coupled with a novel Cut Cell approach to tackle two-phase Stefan problems and an adjoint-based optimization procedure. The key features of the level set related algorithms are, (i) an implicit-explicit scheme to solve the level set advection allowing us to relax the usual CFL condition, (ii) the high order Johansen-Colella method used to compute the normal gradient across the interface, and (iii) the sub-cell resolution reinitialization procedure to retain the signed distance function property as the interface moves. Moreover, we use a novel Cut Cell method coupled with a Crank-Nicolson time integrator that allows us to solve the two-phase problem for any given geometry. The adjoint-based optimization procedure is shown to be a robust algorithm to control the shape of a melting or solidification front, even in the presence of dendritic instabilities and anisotropic effects. As a future work, one could consider to solve the Navier-Stokes equations in the fluid phase, allowing for Rayleigh-Bénard instabilities in the case of the unidirectional solidification process.


\appendix
\section{Adjoint temperature derivation} \label{app:adjoint_derivation}

We detail the derivations from the Lagrange function $\mathcal{L}$ (Eq.~\ref{eq:Lagrange}) to the adjoint Stefan problem  (\ref{AP}). We have the initial conditions,
\begin{equation}
\begin{aligned}
T(x, 0) = T_0(x),\\
\phi(x, 0) = \phi_0(x)
\end{aligned}
\end{equation}
meaning that when calculating the derivatives in the direction $\delta T$ and $\delta \phi$, we require : $\delta T(x, 0) = \delta \phi (x, 0) = 0$.
We consider the adjoint temperature by setting $\mathcal{L}_T(\cdot) = 0$,

\begin{equation}
\begin{aligned}
\mathcal{L}_T \delta T = 
- \displaystyle \int_{\Omega} \beta_1 \left( T(t_f) - T_{t_f}\right) \delta T(t_f)\\
- \displaystyle \int_{0}^{t_{f}} \int_{\Omega_1} \left( \dfrac{\partial \delta T_{1}}{\partial t} - \Delta \delta T_{1} \right) \theta_1 - \int_{0}^{t_{f}} \int_{\Omega_2} \left( \dfrac{\partial \delta T_{2}}{\partial t} - \Delta \delta T_{2} \right) \theta_2 \\
- \displaystyle \int_{0}^{t_{f}} \int_{\partial \Omega} \delta T  \theta_D
- \displaystyle \int_{0}^{t_{f}} \int_\Gamma \delta T  \theta_I \\
- \displaystyle \int_{0}^{t_{f}} \int_\Gamma  \psi [\nabla \delta T_i]^1_2 \cdot \nabla \phi 
\end{aligned}
\end{equation}

We move the spatial and temporal derivatives towards the adjoint state $\theta$. We apply integration by parts, once with respect to time (Corollary \ref{coro:reynolds}) and twice with respect to space (using Green's formula),

\begin{equation}
\begin{aligned}
\mathcal{L}_T \delta T = 
- \displaystyle \int_{\Omega} \beta_1 \left( T(t_f) - T_{t_f}\right) \delta T(t_f)\\
-\displaystyle \int_{\Omega_1(t_f)} \delta T(t_f) \theta_1 (t_f) + \int_{\Omega_1(0)} \delta T(0) \theta_1 (0) \\
- \displaystyle \int_{0}^{t_{f}} \int_{\Omega_1} \dfrac{\partial \theta_1}{\partial t} \delta T + \displaystyle \int_{0}^{t_{f}} \int_{\partial \Omega_1} \delta T \theta_1 V_1\\
+ \displaystyle \int_{0}^{t_{f}} \int_{\partial \Omega_1} \dfrac{\partial \delta T}{\partial n} \theta_1 - \displaystyle \int_{0}^{t_{f}} \int_{\Omega_1} \nabla \delta T \nabla \theta_1\\
-\displaystyle \int_{\Omega_2(t_f)} \delta T(t_f) \theta_2 (t_f) + \int_{\Omega_2(0)} \delta T(0) \theta_2 (0) \\
- \displaystyle \int_{0}^{t_{f}} \int_{\Omega_2} \dfrac{\partial \theta_2}{\partial t} \delta T + \displaystyle \int_{0}^{t_{f}} \int_{\partial \Omega_2} \delta T \theta_2 V_2\\
+ \displaystyle \int_{0}^{t_{f}} \int_{\partial \Omega_2} \dfrac{\partial \delta T}{\partial n} \theta_2 - \displaystyle \int_{0}^{t_{f}} \int_{\Omega_2} \nabla \delta T \nabla \theta_2\\
- \displaystyle \int_{0}^{t_{f}} \int_{\partial \Omega} \delta T  \theta_D
- \displaystyle \int_{0}^{t_{f}} \int_\Gamma \delta T  \theta_I \\
- \displaystyle \int_{0}^{t_{f}} \int_\Gamma  \psi [\nabla \delta T_i]^1_2 \cdot \nabla \phi 
\end{aligned}
\end{equation}

with $V_1$ and $V_2$ the velocities of the control volumes $\Omega_1$ and $\Omega_2$ respectively with $V_1 = \vec{V} \cdot n$ and $V_2 = - \vec{V} \cdot n$. These two terms are only non-zero on $\Gamma$. Sorting the terms by their domain of integration and setting $\delta T(0) = 0$, we obtain,

\begin{equation}
\begin{aligned}
\mathcal{L}_T \delta T = 0 = \displaystyle \int_{0}^{t_{f}} \int_{\Omega_1} \left( \dfrac{\partial \theta_1}{\partial t} + \Delta \theta_1 \right) \delta T \\
+ \displaystyle \int_{0}^{t_{f}} \int_{\Omega_2} \left( \dfrac{\partial \theta_2}{\partial t} + \Delta \theta_2 \right) \delta T \\
- \displaystyle \int_{\Omega_ 1} \left(\theta_1 (t_f) - \beta_1 \left( T(t_f) - T_{t_f}\right) \right) \delta T(t_f)
\\
- \displaystyle \int_{\Omega_ 2} \left(\theta_2 (t_f) - \beta_1 \left( T(t_f) - T_{t_f}\right) \right) \delta T(t_f) \\
+ \displaystyle \int_{0}^{t_{f}} \int_{\partial \Omega} - \dfrac{\partial \theta}{\partial n} \delta T + \dfrac{\partial \delta T}{\partial n} \left(\theta - \theta_D\right) \\
+ \displaystyle \int_{0}^{t_{f}} \int_{\Gamma} \left(\theta_1 \vec{V} \cdot n - \theta_2 \vec{V} \cdot n - \dfrac{\partial \theta_1}{\partial n} + \dfrac{\partial \theta_2}{\partial n} - \theta_I \right) \delta T \\
+ \displaystyle \int_{0}^{t_{f}} \int_{\Gamma} \left( \theta - \psi |\nabla \phi|\right) [\nabla \delta T]^1_2 \cdot n
\end{aligned}
\end{equation}
with the second to last term simplifying to,
$\theta_I = -[\nabla \theta]^1_2 \cdot n$ as $\theta_1 = \theta_2$ on $\Gamma$. 

\begin{thm}[Reynolds transport theorem]\label{theo:reynolds}

The derivative of the quantity

$$
F(t):=\int_{\Omega(t)} f(x, t) d x
$$
is given by
$$
\begin{aligned}
\frac{d F}{d t}(t) &=\int_{\Omega(t)} \frac{\partial f}{\partial t}+\operatorname{div}(f V) d x \\
&=\int_{\Omega(t)} \frac{d f}{d t}+f \operatorname{div}(V) d x \\
&=\int_{\Omega(t)} \frac{\partial f}{\partial t} d x+\int_{\partial \Omega(t)} f V \cdot n d s
\end{aligned}
$$
where $V$ is the velocity field in which the control volume $\Omega(t)$ moves.
\end{thm}

\begin{lem}[Integration by Parts in Time in Moving Domains]\label{coro:reynolds} 
For $g=$ $g(x, t)$ and $h=h(x, t)$, we have
$$
\begin{array}{c}
\displaystyle \int_{0}^{t_f} \int_{\Omega(t)} g h_{t} d x d t=\int_{\Omega(t_f)} g(x, t_f) h(x, t_f) d x-\int_{\Omega(0)} g(x, 0) h(x, 0) d x \\
-\displaystyle \int_{0}^{t_f} \int_{\Omega(t)} g_{t} h d x d t-\int_{0}^{t_f} \int_{\partial \Omega(t)} g h V \cdot n d s d t
\end{array}
$$
\end{lem}

\bibliography{biblist}

\begin{thebibliography}{10}
\expandafter\ifx\csname url\endcsname\relax
  \def\url#1{\texttt{#1}}\fi
\expandafter\ifx\csname urlprefix\endcsname\relax\def\urlprefix{URL }\fi
\expandafter\ifx\csname href\endcsname\relax
  \def\href#1#2{#2} \def\path#1{#1}\fi

\bibitem{sarler_stefans_1995}
B.~Šarler, Stefan's work on solid-liquid phase changes, Engineering Analysis
  with Boundary Elements 16~(2) (1995) 83--92.
\newblock \href {https://doi.org/10.1016/0955-7997(95)00047-X}
  {\path{doi:10.1016/0955-7997(95)00047-X}}.

\bibitem{osher_fronts_1988}
S.~Osher, J.~Sethian, Fronts propagating with curvature-dependent speed:
  {Algorithms} based on {Hamilton}-{Jacobi} formulations, J Comput. Phys.
  79~(1) (1988) 12--49.
\newblock \href {https://doi.org/10.1016/0021-9991(88)90002-2}
  {\path{doi:10.1016/0021-9991(88)90002-2}}.

\bibitem{Juric1996}
D.~Juric, G.~Tryggvason, A front-tracking method for dendritic solidification,
  J Comput. Phys. 123 (1996) 127--148.

\bibitem{Segal1998}
G.~Segal, C.~Vuik, F.~Vermolen, A conserving discretization for the free
  boundary in a two-dimensional stefan problem, J Comput. Phys. 141 (1998)
  1--21.

\bibitem{Hassan2021}
A.~Hassan, T.~Sayadi, V.~LeChenadec, H.~Pitsch, A.~Attili, Adjoint-based
  sensitivity analysis of steady char burn out, Com. Theo. Modeling 25~(1)
  (2021) 96--120.
\newblock \href {https://doi.org/10.1080/13647830.2020.1838614}
  {\path{doi:10.1080/13647830.2020.1838614}}.

\bibitem{javierre_comparison_2006}
E.~Javierre, C.~Vuik, F.~Vermolen, S.~van~der Zwaag, A comparison of numerical
  models for one-dimensional {Stefan} problems, J Comp. Applied Mathematics
  192~(2) (2006) 445--459.
\newblock \href {https://doi.org/10.1016/j.cam.2005.04.062}
  {\path{doi:10.1016/j.cam.2005.04.062}}.

\bibitem{Rose1993}
M.~Rose, An enthalpy scheme for stefan problems in several dimensions, App.
  Numerical Math. 12 (193) 229.

\bibitem{Brattkus1992}
K.~Brattkus, D.~Meiron, Numerical simulations of unsteady crystal growth, SIAM
  J. Appl. Math. 52 (1992) 1303.

\bibitem{Limare2022}
A.~Limare, S.~Popinet, C.~Josserand, Z.~Xue, A.~Ghigo, A hybrid level-set /
  embedded boundary method applied to solidification-melt problems, arXiv
  (2022).
\newblock \href {http://arxiv.org/abs/2202.08300} {\path{arXiv:2202.08300}}.

\bibitem{langer_instabilities_1980}
J.~S. Langer, Instabilities and pattern formation in crystal growth, Rev. Mod.
  Phys. 52~(1) (1980) 1--28.
\newblock \href {https://doi.org/10.1103/RevModPhys.52.1}
  {\path{doi:10.1103/RevModPhys.52.1}}.

\bibitem{mullins_stability_1964}
W.~W. Mullins, R.~F. Sekerka,
  \href{https://aip.scitation.org/doi/10.1063/1.1713333}{Stability of a planar
  interface during solidification of a dilute binary alloy}, Journal of Applied
  Physics 35~(2) (1964) 444--451, publisher: American Institute of Physics.
\newblock \href {https://doi.org/10.1063/1.1713333}
  {\path{doi:10.1063/1.1713333}}.
\newline\urlprefix\url{https://aip.scitation.org/doi/10.1063/1.1713333}

\bibitem{woods_melting_1992}
A.~Woods, Melting and dissolving, J. Fluid Mech. 239 (1992) 429.
\newblock \href {https://doi.org/10.1017/S0022112092004476}
  {\path{doi:10.1017/S0022112092004476}}.

\bibitem{Marsden2008}
A.~L. Marsden, J.~A. Feinstein, C.~A. Taylor, {A computational framework for
  derivative-free optimization of cardiovascular geometries}, Computer Methods
  in Applied Mechanics and Engineering 197~(21-24) (2008) 1890--1905.

\bibitem{Pierret2007}
S.~Pierret, R.~F. Coelho, H.~Kato, {Multidisciplinary and multiple operating
  points shape optimization of three-dimensional compressor blades}, Structural
  and Multidisciplinary Optimization 33~(1) (2007) 61--70.

\bibitem{Giles2000}
M.~B. Giles, N.~A. Pierce, {An Introduction to the Adjoint Approach to Design},
  Flow Turb. Combus. 65~(3) (2000) 393--415.

\bibitem{Pironneau1974}
O.~Pironneau, {On optimum design in fluid mechanics}, J Fluid Mech. 64~(1)
  (1974) 97--110.

\bibitem{Jameson88}
A.~Jameson, {Aerodynamic design via control theory}, J Scientific Computing
  3~(3) (1988) 233--260.

\bibitem{Jameson98}
A.~Jameson, L.~Martinelli, N.~A. Pierce, {Optimum Aerodynamic Design Using the
  Navier-Stokes Equations}, Theor. Comp. Fluid Dyn. 10~(1) (1998) 213--237.

\bibitem{Juniper2010}
M.~Juniper, {Triggering in the horizontal Rijke tube: non-normality, transient
  growth and bypass transition}, J. Fluid Mech. 667 (2010) 272--308.

\bibitem{Lemke2013}
M.~Lemke, J.~Reiss, J.~Sesterhenn, {Adjoint-based analysis of thermoacoustic
  coupling}, ICNAAM (2013) 2163--2166.

\bibitem{Schmidt2013}
S.~Schmidt, C.~Ilic, V.~Schulz, N.~R. Gauger, Three-dimensional large-scale
  aerodynamic shape optimization based on shape calculus, AIAA J 51~(11) (2013)
  2615--2627.

\bibitem{Rabin2014}
S.~Rabin, C.~Caulfield, R.~Kerswell, {Designing a more nonlinearly stable
  laminar flow via boundary manipulation}, J Fluid Mech. 738 (2014) R1:1--12.

\bibitem{Foures2014}
D.~Foures, C.~Caulfield, P.~Schmid, {Optimal mixing in two-dimensional plane
  Poiseuille flow at finite Péclet number}, J Fluid Mech. 748 (2014) 241--277.

\bibitem{Duraisamy2012}
K.~Duraisamy, J.~Alonso, Adjoint-based techniques for uncertainty
  quantification in turbulent flows with combustion, 42nd AIAA Fluid Dynamics
  Conference and Exhibit (2012) 25--28.

\bibitem{Fikl2020}
A.~Fikl, V.~{Le Chenadec}, T.~Sayadi, Control and optimization of interfacial
  flows using adjoint-based techniques, Fluids 5~(3) (2020).
\newblock \href {https://doi.org/10.3390/fluids5030156}
  {\path{doi:10.3390/fluids5030156}}.

\bibitem{ou_unsteady_2011}
K.~Ou, A.~Jameson, \href{https://arc.aiaa.org/doi/10.2514/6.2011-24}{Unsteady
  adjoint method for the optimal control of advection and burger's equations
  using high-order spectral difference method}, American Institute of
  Aeronautics and Astronautics (2011) 1--18\href
  {https://doi.org/10.2514/6.2011-24} {\path{doi:10.2514/6.2011-24}}.
\newline\urlprefix\url{https://arc.aiaa.org/doi/10.2514/6.2011-24}

\bibitem{Braman2015}
K.~Braman, T.~Oliver, V.~Raman, Adjoint-based sensitivity analysis of flames,
  Comb. Theo. Modelling 19~(1) (2015) 29--56.

\bibitem{Lemke2019}
M.~Lemke, L.~Cai, J.~Reiss, H.~Pitsch, J.~Sesterhehn, Adjoint-based sensitivity
  analysis of quantities of interest of complex combustion models, Comb. Theo.
  Modelling 23~(1) (2019) 180--196.

\bibitem{Hoffmann1982}
K.~Hoffmann, J.~Sprekels, Real-time control of the free boundary in a two-phase
  stefan problem, Numerical Functional Analysis and Optimization 5~(1) (1982)
  47--76.

\bibitem{Knabner1985}
P.~Knabner, Control of stefan problems by means of linear-quadratic defect
  minimization, Numerische Mathematik 46~(3) (1985) 429--442.

\bibitem{Kang1995}
S.~Kang, N.~Zabaras, Control of the freezing interface motion in
  two-dimensional solidification processes using the adjoint method,
  International J Numerical Methods in Engineering 38 (1995) 63--80.

\bibitem{Yang1997}
Z.~Yang, The adjoint method for the inverse design of solidification processes
  with convection, PhD thesis, Cornell University (1997).

\bibitem{Hinze2007}
M.~Hinze, S.~Ziegenbalg, Optimal control of the free boundary in a two-phase
  stefan problem, J Comput. Phys. 223~(2) (2007) 657--684.

\bibitem{bernauer_optimal_2011}
M.~Bernauer, R.~Herzog, Optimal {Control} of the {Classical} {Two}-{Phase}
  {Stefan} {Problem} in {Level} {Set} {Formulation}, SIAM J. Sci. Comput.
  33~(1) (2011) 342--363.
\newblock \href {https://doi.org/10.1137/100783327}
  {\path{doi:10.1137/100783327}}.

\bibitem{Langer1980}
J.~S. Langer, Instabilities and pattern formation in crystal growth, Reviews of
  Modern Physics (1980).

\bibitem{chen_simple_1997}
S.~Chen, B.~Merriman, S.~Osher, P.~Smereka, A {Simple} {Level} {Set} {Method}
  for {Solving} {Stefan} {Problems}, J Comp. Phys. 135~(1) (1997) 8--29.
\newblock \href {https://doi.org/10.1006/jcph.1997.5721}
  {\path{doi:10.1006/jcph.1997.5721}}.

\bibitem{juric_front-tracking_1996}
D.~Juric, G.~Tryggvason, A {Front}-{Tracking} {Method} for {Dendritic}
  {Solidification}, J Comp. Phys. 123~(1) (1996) 127--148.
\newblock \href {https://doi.org/10.1006/jcph.1996.0011}
  {\path{doi:10.1006/jcph.1996.0011}}.

\bibitem{FullanaCutCell}
T.~Fullana, A.~Q. Rodr{\'i}guez, V.~L. Chenadec, T.~Sayadi, A cut cell method
  for the solution of the incompressible navier-stokes equations around
  stationnary geometries, Journal of Computational Physics (2022).

\bibitem{johansen_cartesian_1998}
H.~Johansen, P.~Colella, A cartesian grid embedded boundary method for
  poisson's equation on irregular domains, J Comp. Phys. 147~(1) (1998) 60--85.
\newblock \href {https://doi.org/10.1006/jcph.1998.5965}
  {\path{doi:10.1006/jcph.1998.5965}}.

\bibitem{peng_pde-based_1999}
D.~Peng, B.~Merriman, S.~Osher, H.~Zhao, M.~Kang, A {PDE}-{Based} {Fast}
  {Local} {Level} {Set} {Method}, J Comp. Phys. 155~(2) (1999) 410--438.
\newblock \href {https://doi.org/10.1006/jcph.1999.6345}
  {\path{doi:10.1006/jcph.1999.6345}}.

\bibitem{mikula_inflow-implicitoutflow-explicit_2014}
K.~Mikula, M.~Ohlberger, J.~Urbán, Inflow-implicit/outflow-explicit finite
  volume methods for solving advection equations, Applied Numerical Mathematics
  85 (2014) 16--37.
\newblock \href {https://doi.org/10.1016/j.apnum.2014.06.002}
  {\path{doi:10.1016/j.apnum.2014.06.002}}.

\bibitem{mikula_new_2010}
K.~Mikula, M.~Ohlberger, A {New} {Level} {Set} {Method} for {Motion} in
  {Normal} {Direction} {Based} on a {Semi}-{Implicit} {Forward}-{Backward}
  {Diffusion} {Approach}, SIAM J. Sci. Comput. 32~(3) (2010) 1527--1544.
\newblock \href {https://doi.org/10.1137/09075946X}
  {\path{doi:10.1137/09075946X}}.

\bibitem{min_reinitializing_2010}
C.~Min, On reinitializing level set functions, J Comp. Phys. 229~(8) (2010)
  2764--2772.
\newblock \href {https://doi.org/10.1016/j.jcp.2009.12.032}
  {\path{doi:10.1016/j.jcp.2009.12.032}}.

\bibitem{popinet_accurate_2009}
S.~Popinet,
  \href{https://linkinghub.elsevier.com/retrieve/pii/S002199910900240X}{An
  accurate adaptive solver for surface-tension-driven interfacial flows},
  Journal of Computational Physics 228~(16) (2009) 5838--5866.
\newblock \href {https://doi.org/10.1016/j.jcp.2009.04.042}
  {\path{doi:10.1016/j.jcp.2009.04.042}}.
\newline\urlprefix\url{https://linkinghub.elsevier.com/retrieve/pii/S002199910900240X}

\bibitem{Frank1950}
F.~C. Frank, Radially symmetric phase growth controlled by diffusion, Proc. R.
  Soc. Lond (1950).
\newblock \href {https://doi.org/http://doi.org/10.1098/rspa.1950.0080}
  {\path{doi:http://doi.org/10.1098/rspa.1950.0080}}.

\bibitem{ALMGREN1993}
R.~Almgren, Variational algorithms and pattern formation in dendritic
  solidification, J Comp. Phys. 106~(2) (1993) 337--354.
\newblock \href {https://doi.org/https://doi.org/10.1016/S0021-9991(83)71112-5}
  {\path{doi:https://doi.org/10.1016/S0021-9991(83)71112-5}}.

\bibitem{Tan2006}
L.~Tan, N.~Zabaras, A level set simulation of dendritic solidification with
  combined features of front-tracking and fixed-domain methods, Journal of
  Computational Physics 211~(1) (2006) 36--63.
\newblock \href {https://doi.org/10.1016/j.jcp.2005.05.013}
  {\path{doi:10.1016/j.jcp.2005.05.013}}.

\bibitem{Liu1989}
D.~C. Liu., J.~Nocedal, On the limited memory method for large scale
  optimization, Mathematical Programming 3~(45) (1989) 503–528.

\bibitem{Wright2006}
S.~J. Wright, J.~Nocedal, Numerical optimization, Springer, 2006.

\bibitem{julia2015}
J.~Bezanson, A.~Edelman, S.~Karpinski, V.~B. Shah, Julia: A fresh approach to
  numerical computing, SIAM Review 59~(1) (2017) 65--98.
\newblock \href {https://doi.org/10.1137/141000671}
  {\path{doi:10.1137/141000671}}.

\bibitem{Mogensen2018}
P.~K. Mogensen, A.~N. Riseth, Optim: A mathematical optimization package for
  {Julia}, Journal of Open Source Software 3~(24) (2018) 615.
\newblock \href {https://doi.org/10.21105/joss.00615}
  {\path{doi:10.21105/joss.00615}}.

\end{thebibliography}

\end{document}